\documentclass[twocolumn,preprintnumbers,showpacs,amsmath,amssymb,floatfix]{revtex4}
\def\address{\affiliation}
\usepackage{graphicx,array,subfigure}
\newcommand{\partwo}[2]{\ensuremath{\frac{\partial^2 #1}{\partial #2^2}}}
\newcommand{\partwomix}[3]{\ensuremath{\frac{\partial^2 #1}{{\partial #2}{\partial #3}}}}
\newcommand{\parone}[2]{\ensuremath{\frac{\partial #1}{\partial #2}}}
\newcommand{\textparone}[2]{\ensuremath{{\partial #1}/{\partial #2}}}

\newcommand{\coss}[2]{\ensuremath{\cos  #1 #2 }}
\newcommand{\tm}{\ensuremath{\tilde m}}
\newcommand{\tn}{\ensuremath{\tilde n}}
\newcommand{\tp}{\ensuremath{\tilde p}}
\newcommand{\zeff}{\ensuremath{z_{\mathrm{eff}}}}
\newcommand{\bico}{$(\mathrm{Bi}_{0.8}\mathrm{Pb}_{0.2})_2\mathrm{Sr}_2\mathrm{Co}_2\mathrm{O}_\mathrm{y}$}
\begin{document}

\title{
Thermoelectric Effects in Anisotropic Systems: Measurement and Applications.
}

\author{
T. W. Silk and A. J. Schofield}

\address{
School of Physics and Astronomy, The University of Birmingham, Birmingham, B15 2TT, U.K.
}

\begin{abstract}

The Harman method for measuring the thermal conductivity of a sample using the Peltier effect, may also be used to determine the dimensionless figure of merit from just two electrical resistance measurements.  We consider a modified version of the Harman method where the current contacts are much smaller than the contact faces of the sample.  We calculate the voltage and temperature distributions in a rectangular sample of a material having anisotropy in all of its transport coefficients.  The thermoelectric anisotropy has important consequences in the form of thermoelectric eddy currents and the Bridgman effect.  We prove that in the limit of a very thin sample of arbitrary shape, there exist van der Pauw formulae relating particular linear combinations of the potential and temperature differences between points on the edges of the sample.  We show that the Harman figure of merit can be radically different from the intrinsic figures of merit of the material, and can often be substantially enhanced.  By defining an effective figure of merit in terms of the rate of entropy production, we show that the increase in the Harman figure of merit does indicate an improvement in the thermoelectric performance of an anisotropic sample having small current contacts.  However, we also prove that in the case of a material with tetragonal symmetry, the effective figure of merit is always bounded from above by the largest intrinsic figure of merit of the material.    

\end{abstract}

\pacs{44.10.+i, 72.15.Jf}

\maketitle

\section{Introduction} \label{sec:Introduction}

It is widely accepted that improving the efficiency of thermoelectric devices is one of the key challenges currently facing both technologists and physicists alike.  A thermoelectric device either converts heat into electrical power through the Seebeck effect, or employs the Peltier effect to transport heat.  The efficiency of such devices depends primarily on the transport coefficients of the constituent thermoelectric materials.  The ideal thermoelectric material would have i) a large thermopower $S$ to give maximum transfer of heat into electrical power or vice versa; ii) a large electrical conductivity $\sigma$ to minimise the wasted power associated with Joule heating; and iii) a low thermal conductivity $\kappa$ to minimise thermal shorting.  The thermal conductivity will, in general, include both contributions from the lattice, and from the charge carriers themselves.  The actual efficiency of a thermoelectric device will generally depend in a very complicated way on the transport coefficients of its composite materials \cite{Sherman}.  However, a simple result is obtained in the case of a single thermoelectric material operating as a generator between two temperatures, and whose transport coefficients are taken to be temperature independent.  The maximum efficiency of such a device turns out to depend on the transport coefficients only through a quantity known as the figure of merit (FOM), $Z = S^2 \sigma/\kappa$.  The larger the value of $Z$, the higher the efficiency.  The focus is therefore on discovering or designing materials with large figures of merit.  A dimensionless FOM, $z=ZT$ in excess of unity is needed to make thermoelectric devices competitive in terms of efficiency \cite{Mahan2}.

The ability to accurately determine the FOM is of great importance.  One way to determine the FOM is simply to make separate measurements of all three transport coefficients individually.  However, there is a commonly employed technique which allows a determination of the dimensionless FOM from just two electrical resistance measurements.  This is known as the Harman method \cite{Harman}, and was originally designed as a means of measuring the thermal conductivity using the Peltier effect.  An electrical current passed through the sample results in a temperature difference across the sample due to the Peltier effects occurring at each contact.  The steady state temperature difference results from the balance between the Peltier effects and the heat flow by thermal conduction, which allows the thermal conductivity to be determined.  It turns out that the electrical resistances of the sample due to alternating and direct current in the Harman set-up allow the dimensionless FOM of the sample to be determined from $z = R_{\mathrm{dc}}/R_{\mathrm{ac}} -1$. This result depends on a large number of assumptions, mainly relating to the various heat losses  and heating effects which can occur in the sample.  However, there is an additional assumption, which is that the flows of heat and electricity are uniform through the sample.  It is important to ask what would happen if this assumption were not fulfilled.  Such a situation might occur if the contacts between the sample and the current leads are smaller than the cross-sectional area of the sample itself.

We will show later in this work that the effects of the small current contact do not change the FOM measured with the Harman method provided that the transport coefficients of the sample are isotropic.  However, it is extremely unlikely that this would occur in practice.  Indeed, many of the materials which are currently under investigation as potential thermoelectrics have anisotropic transport properties.  These include the layered cobalt oxides \cite{Terasaki1}, and semiconductor superlattices \cite{Venk, Harman2, Yang}.  Even traditional materials such as $\mathrm{Bi}_2\mathrm{Te}_3$ have uniaxially anisotropic transport properties \cite{Rowe}.  In particular, anisotropy in the thermopower can lead to new and interesting effects not seen in isotropic systems.  First, there is an additional internal heating effect, the Bridgman effect \cite{Bridgman, Domenicalli}, which occurs if the current distribution becomes non-uniform.  Importantly, this effect is linear in the electrical current, and hence cannot be neglected as a higher-order process.  Second, the steady state with an applied temperature distribution generally contains circulating electrical currents, known as thermoelectric eddy currents \cite{Samoilovich}.  These lead to a non-trivial relationship between the voltage and temperature distributions in the sample.  
     
The theory of electrical resistance measurements with small current and voltage contacts is extremely well-established in the literature.  The classic work of van der Pauw \cite{vdp1} considered electrical resistance measurements in thin samples of arbitrary shape.  It was shown that two particular  four-terminal resistance measurements on an isotropic sample obey a universal relationship, allowing the electrical resistivity to be determined.  Logan \textit{et al} \cite{Logan} considered the problem of determining the voltage distribution in a rectangular sample of isotropic resistivity with current contacts on the corners.  Their solution was subsequently used by Montgomery \cite{Montgomery} to show how all three components of the resistivity tensor in an anisotropic sample can be determined from appropriate resistance measurements.  To our knowledge, the equivalent calculations have not been extended to thermal and thermoelectric measurements.    

A motivation for this work is recent experimental work \cite{Kobayashi, Tamura} where the Harman method is used to measure the FOM of various anisotropic materials.  The contact configurations used in the experiments are very similar to those used in Montgomery's methods of measuring the electrical resistivity, with small current and voltage contacts on the surface of the sample.  In some configurations, the Harman FOM (\textit{i.e.} the FOM measured using the Harman method) is in good agreement with the known intrinsic FOM of the material.  However, in measurements on \bico \ (BiPbSrCoO) single crystals in the out-of-plane direction, the Harman FOM is found to be substantially enhanced above the intrinsic FOM.  A key feature of this material is that while the thermal conductivity is relatively isotropic, the electrical conductivity exhibits an anisotropy of around $10^4$ between in and out-of-plane directions.  We will discuss the results of these experiments in some detail later in the paper.  We also note that other experiments on $(\mathrm{Bi}, \mathrm{Sb})_2\mathrm{Te}_3$ \cite{Iwasaki} have also revealed an intriguing dependence of the Harman FOM on the sample length, cross-sectional area and voltage contact location.    

The principle aim of this paper is to calculate the voltage and temperature distributions in a sample with small current contacts, and having arbitrary degrees of anisotropy in its transport coefficients.  We will then investigate how this alters the FOM measured in the Harman method.  We will show that the Harman FOM no-longer takes a single value for the sample, and is in general a function of the points between the resistances are measured.  In some configurations the Harman FOM is substantially enhanced above the intrinsic FOM of the material in the measurement direction.  By defining an effective FOM based on the rates of entropy production in the Harman method, we show that the increased Harman FOM does indicate an improvement in the thermoelectric performance of the system.  We identify two important mechanisms responsible for this increase.  First, in a system whose electrical and thermal anisotropies are different, heat and electricity can flow along very different paths.  Second, thermoelectric eddy currents and the Bridgman effect lead to a mixing of the thermopowers in the different coordinate directions, and can lead to an increase in the size of the thermoelectric effects.  We will also prove, however, that the effective FOM is always bounded from above by the largest intrinsic FOM in an anisotropic material.         

The rest of this paper is organised as follows.  In section \ref{sec:theory}, we set-out the theory needed to describe transport in anisotropic systems.  In particular, we will discuss the origins of the new thermoelectric phenomena, namely eddy currents and the Bridgman effect.  In section \ref{sec:Harman} we describe the Harman method in some detail, discuss the various assumptions made, and show how some of the errors which occur can be reduced.  In section \ref{sec:model} we present the model, and obtain the most general solution.  The solution can be simplified once the current contacts locations are fixed, and we derive the simplified solutions for two different configurations in appendices \ref{sec:same} and \ref{sec:opp}.  The Harman FOM is calculated in section \ref{sec:hfom} for several different cases of anisotropy and sample geometry.  In addition, we consider the limit of a very thin sample, and show that equivalents of van der Pauw's relation hold in the Harman method.  In section \ref{sec:case} we apply the results to the recent experiments of Kobayashi \textit{et al} and Tamura \textit{et al}. Finally, in sections \ref{sec:efffom} and \ref{sec:bounds} we investigate whether the increased Harman FOM indicates an improvement in thermoelectric performance.  This leads us to a new definition of the FOM in terms of the rates of entropy production in the Harman method.        

\section{Theory of Anisotropic Transport}\label{sec:theory}

In an anisotropic system, the $i$th components of the electrical and heat current densities, $\mathbf j_i$ and $\mathbf j_Q$, are given respectively by the linear response equations $(x_i = \{x,y,z\})$
\begin{align}
\label{eq:anisoelec}
&\j_i = \sigma_{ik}\parone{\chi}{x_k} - \sigma_{il}S_{lk}\parone{T}{x_k}, \\
\label{eq:anisoheat}
&\j_{Qi} = \Pi_{ik}j_{k} - \kappa_{ik}\parone{T}{x_k},
\end{align}
where the summation convention has been invoked. Here, $T \equiv T(x,y,z)$ is the absolute temperature, and $\chi \equiv \chi(x,y,z)$ is the electrochemical potential.  This is defined as $\chi = \mu/e - \phi$, where $\mu$ is the chemical potential, $\phi$ is the electrical potential, and $e$ is the magnitude of the electronic charge.  The Kelvin/Onsager relations imply that the Peltier and Seebeck coefficients are related by $\Pi_{ik} = TS_{ki}$.  In this work, all of the transport coefficients are taken to be diagonal tensors.  Effects associated with the off-diagonal components \cite{HarmanHonig, Bies} will not be considered here. 

Let us now consider the Seebeck effect in an anisotropic system.  In the conventional picture of a homogeneous, isotropic system, a temperature gradient applied to an isolated sample causes an electrical current to flow, until the build-up of charge creates a voltage gradient which counteracts it.  This leads to a steady state in which no net electrical current flows.  Setting $j_i=0$ and $S_{ik} = S\delta_{ik}$ in equation (\ref{eq:anisoelec}), and solving for $\chi$ gives the familiar result that the thermoelectric voltage is given by (up to an arbitrary constant of integration) $\chi_{\mathrm{th}} = ST$ in the steady state.  

When the thermopower is anisotropic, the steady state does not have $j_i=0$ in general.  To see this we first rearrange equation (\ref{eq:anisoelec}) for $\textparone{\chi}{x_i}$
\begin{equation}\label{eq:anisochi}
\parone{\chi}{x_i} = \rho_{ik} j_k + S_{ik}\parone{T}{x_k},
\end{equation}
where $\rho = \sigma^{-1}$ is the resistivity tensor.  If $j_i=0$ is a solution of this equation, then we should be able to integrate to find $\chi$ in terms of $T$.  This is only possible if $\nabla \chi$ is a conservative vector field, and hence $\nabla \times (\nabla \chi) =0$.  Setting $j_i=0$ and requiring that the curl of the right-hand side of equation (\ref{eq:anisochi}) equals zero implies that
\begin{equation}\label{eq:bridgcond}
(S_x-S_y)\partwomix{T}{x}{y} =0 = (S_x-S_z)\partwomix{T}{x}{z} = (S_y-S_z)\partwomix{T}{y}{z} \nonumber \\,
\end{equation}
where $S_x \equiv S_{xx}$ \textit{etc}.  Hence unless the temperature distribution depends on only one coordinate, we see that the steady state where all components of the current density vanish cannot exist when the thermopower is anisotropic.  This implies that in the steady state there are electrical currents flowing in the bulk, providing an Ohmic component to the voltage which ensures that $\nabla \times (\nabla \chi) =0$.  These currents, often referred to as \textit{thermoelectric eddy currents}, were first predicted and observed by Samoilovich and Korenblit \cite{Samoilovich}.  We note that their existence has the following important consequence;  \textit{the relationship between the thermoelectric voltage and temperature distributions does not generally obey $\chi_{\mathrm{th}} = ST$ when the thermopower is anisotropic, even if $S$ itself is independent of temperature.}  

We now turn to the internal heating effects which can occur in an anisotropic system.  These effects may be derived by applying the principles of conservation of charge and conservation of energy \cite{Domenicalli}, which can be expressed in terms of continuity equations $\textparone{\rho}{t} + \nabla . \mathbf{j}=0$ and $\textparone{u}{t} + \nabla . \mathbf{j}_U=0$.  Here, $\rho$ and $u$ are the charge and energy densities respectively.  The quantity $\mathbf{j}_U$ is the energy current density, related to the heat current density via $\mathbf j_U = \mathbf j_Q - \chi \mathbf j$. In a steady state, the continuity equations become
\begin{eqnarray}\label{eq:eleccont}
&\displaystyle{\parone{j_i}{x_i} = 0}, \\
&\displaystyle{\parone{j_{Ui}}{x_i} = \parone{}{x_i}\Big((TS_{ik} - \chi \delta_{ik})j_k - \kappa_{ik}\parone{T}{x_k}\Big) = 0}.
\end{eqnarray}    
Expanding out the right-hand side of the second of these equations and making use of equation (\ref{eq:anisoelec}) leads to the following equation:
\begin{multline}\label{eq:heatcont} 
 -\parone{}{x_i}\Big ( \kappa_{ik}\parone{T}{x_k}\Big) = \rho_{ik}j_i j_k \\
-Tj_k\Big(\parone{S_{ik}}{x_i}\Big )_T -\Big(T\parone{S_{ik}}{T}\Big)j_k\parone{T}{x_i} -TS_{ik}\parone{j_k}{x_i}.
\end{multline}
The left-hand side is the divergence of the heat current which would occur in the absence of an electrical current.  The terms on the right-hand side are source terms which represent the internal heating effects.  The first term is recognised as the Joule heating, while the remaining terms contain the thermoelectric effects.  The first two terms on the second line are the (internal) Peltier effect and the Thomson effect, both of which are familiar from isotropic systems.  These terms originate from the fact that if the thermopower varies with position, then as an electrical current flows through the sample the amount of heat it carries also changes.  This variation can come from an intrinsic inhomogeneity of the sample (Peltier effect), or because the thermopower is a function of temperature which itself varies throughout the sample (Thomson effect).  The most common manifestation of the Peltier effect occurs at the junction between two \textit{different} materials, where the rate of heating at the junction is given by the difference of the Peltier coefficients in each material multiplied by the current, $\dot Q = (\Pi_2-\Pi_1)I$.  If the thermopower is anisotropic, then we must also take into account the direction in which the current enters the sample, since this will determine which component of the Peltier tensor appears in this expression.   

The third term on the second line represents an effect originally proposed by Bridgman \cite{Bridgman}.  Unlike the Peltier and Thomson effects, the Bridgman effect only occurs when the thermopower is anisotropic.  For isotropic thermopower $S_{ik} = S\delta_{ik}$, and the Bridgman term reduces to $-TS\textparone{j_i}{x_i}$, which vanishes in the steady state.  To understand the physical origin of this term, consider a current flowing in the $x$-direction (say) in an anisotropic material.  It carries an amount of heat proportional to $\Pi_x$.  If the current were to change direction, say into the $y$-direction, it would now carry a \textit{different} amount of heat, now proportional to $\Pi_y$.  The Bridgman effect therefore captures the heating and cooling effects associated with changes in the direction of the current in a system with an anisotropic thermopower.

Having outlined the basic theory we shall require in this paper, we now turn to a discussion of the conventional Harman method.

\section{The Harman Method}\label{sec:Harman}

In the 1950's, Harman \cite{Harman} proposed a novel technique for measuring the thermal conductivity of a material, which utilises thermoelectric effects.  In the conventional set-up, current leads are attached to the sample in such a way that the electrical current flows in the direction in which the thermal conductivity is to be measured, the $z$-direction say.  The leads are then used to suspend the sample in a vacuum.  As the current enters sample, the Peltier effect causes a heating/cooling of the contact area, with an equal cooling/heating as it leaves through the other contact.  The rate of heating/cooling is given by $(\Pi_l - \Pi_z)I$, where $\Pi_l$ and $\Pi_z$ are the Peltier coefficients of the leads and the sample (in the $z$-direction).  The temperature gradient created in the sample results in a back-flow of heat by thermal conduction, which eventually balances the Peltier effects leading to a steady state.  The simplest treatment of this problem makes the following assumptions; (i) All internal heating effects (as described in section \ref{sec:theory}) in the sample may be neglected; (ii) All heat losses from the sample may also be neglected; (iii) The electrical current is injected/removed uniformly across the contact faces of the sample. Hence, the Peltier effects cause the contact faces to be heated/cooled to uniform temperatures $T_h$/$T_c$, and the flows of heat and electricity may be treated as being effectively one-dimensional.

Under these assumptions, it is straightforward to solve equation (\ref{eq:heatcont}) for the temperature distribution in the sample,
\begin{equation}\label{eq:1dtemp}
T(x,y,z) = \bar T - \frac{(\Pi_l - \Pi_z) Ic}{\kappa_z ab}\Big ( \frac{z}{c} -\frac{1}{2} \Big ),
\end{equation}
where the contact faces of the sample lie at $z=0,c$, and have an area $a \times b$.  The ambient temperature is denoted by $\bar T$.  It follows that the temperature difference between \textit{any} two points $(x,y,0)$, $(x',y',c)$ is 
\begin{align}\label{eq:1ddeltaT}
\Delta T &= T(x',y',c) - T(x,y,0)  \nonumber \\
& = -\frac{(\Pi_l - \Pi_z) Ic}{\kappa_z ab} = -\frac{(\Pi_l - \Pi_z) I}{K_{z0}},
\end{align}
where $K_{z0} = \kappa_z ab/c$ is the thermal conductance of the sample in the $z$-direction.  If $\Pi_l$ and $\Pi_z$ are already known, then measuring $\Delta T$ determines $\kappa_z$.

In addition to measuring the thermal conductivity, the experimental setup of the Harman method may also be used to measure the FOM of the sample directly.  The voltage distribution in the sample with a direct current flowing can be found from equation (\ref{eq:eleccont}):
\begin{equation}\label{eq:1dchi}
\chi(x,y,z) = \bar \chi + \frac{Ic}{\sigma_z ab}\Big ( \frac{z}{c} -\frac{1}{2} \Big ) - \frac{S_z(\Pi_l - \Pi_z) Ic}{\kappa_z ab}\Big ( \frac{z}{c} -\frac{1}{2} \Big ). 
\end{equation}
The potential difference measured across the sample using leads with a thermopower $S_l$ is given by
\begin{equation}\label{eq:1dvolt}   
V_{\mathrm{dc}} = I R_{z0} - (S_l-S_z) \Delta T. 
\end{equation}
In a second measurement, the potential difference is again measured across the sample, but now with an alternating current of amplitude $I$ flowing.  Provided that the period of the current is much shorter than the characteristic time for the temperature difference to develop, there will be no contribution from the thermoelectric term, and hence
\begin{equation}
V_{\mathrm{ac}} = I R_{z0},
\end{equation} 
where $R_{z0} = c/\sigma_z ab$ is the electrical resistance of the sample in the $z$-direction.  Defining a quantity $R_{\mathrm{dc}} = V_{\mathrm{dc}}/I$ and using equation (\ref{eq:1ddeltaT}), the potential difference can be written as
\begin{equation}\label{eq:1dIR1} 
IR_{\mathrm{dc}} = I R_{z0} + (S_l - S_z)\frac{(\Pi_l - \Pi_z) I}{K_{z0}}.
\end{equation}
Using $\Pi_l - \Pi_z = \bar T(S_l - S_z)$, the FOM as determined by the Harman method is then given by
\begin{equation}\label{eq:figofmerit}
z_H = \frac{\bar{T} (S_l-S_z)^2}{R_{z0} K_{z0}} = \frac{V_{\mathrm{dc}} - V_{\mathrm{ac}}}{V_{\mathrm{ac}}} = \frac{R_{\mathrm{dc}} - R_{z0}}{R_{z0}}.  
\end{equation}
Under `normal' circumstances the product $R_{z0} K_{z0} = \rho_z \kappa_z$, since the area and length factors cancel out.  The FOM measured by the Harman method is a relative FOM between the sample and the measuring leads, since the thermopower appearing in (\ref{eq:figofmerit}) is $S_l-S_z$.   The thermopower of the leads is typically much smaller than that of the sample, and so $(S_l-S_z)^2 \approx S_z^2$.  Hence this Harman FOM is approximately equal to the intrinsic FOM of the sample (in the $z$-direction).  

Let us now consider the validity of the assumptions (i)-(iii), which were used to derive the results above.  Assumption (i) is extremely crude, and requires further investigation.  By restricting attention to homogeneous samples, we can eliminate the internal Peltier effects, while the Thomson effect is small and can usually be neglected.  The Bridgman effect is also eliminated provided that assumption (iii) holds, since the current density is uniform throughout the sample.  Note that the 1d nature of the temperature distribution also guarantees that there are no thermoelectric eddy currents in the sample, and that the thermoelectric voltage and temperature distribution are related by $\chi_{\mathrm{th}} = ST$.  A related assumption is that the temperature difference generated is sufficiently small that linear response equations may be applied; this can be achieved by carefully choosing the current.  

The effects of Joule heating are relatively easy to include within the one-dimensional model, since the rate of heat production is constant throughout the system due to the constant current density.  A finite thermal conductance in the leads needs to be included to allow the Joule heat to be removed.  In addition, we include a contact resistance $r_L$ and $r_R$ for each current contact, which may be different.  Again solving equation (\ref{eq:heatcont}), we find that the temperature difference between the contact faces of the sample is now given by   
\begin{equation}\label{eq:quadT}  
\Delta T = T(x',y',c) - T(x,y,0) = \frac{I^2(r_R - r_L)/2- \Pi I}{K_{z0}+K_l/2},
\end{equation}
where $K_l$ is the thermal conductance of a single lead. The result is completely independent of the bulk resistance, and hence we conclude that provided the temperature is measured between the extreme edges of the sample, the bulk Joule heating has no direct effect on $\Delta T$.  The effect of conductive heat loss along the leads reduces the temperature difference across the sample, as one might naively expect.  The current leads will often have a much lower thermal conductance than the sample, and this term can be ignored.  If the contact resistances are not equal, then a convenient way to eliminate their effect is to make measurements of $\Delta T$ for currents passed in both directions through the sample, and calculate $\Delta T^a = (\Delta T(I) - \Delta T(-I))/2$.  This removes the contact resistance contribution, since it is proportional to $I^2$ \cite{Satake}.

We have already dealt with one aspect of assumption (ii) above, namely the heat loss due to thermal conduction along the current leads.  Since the sample is hung in vacuum, convective heat losses are eliminated, leaving radiation as the only other possible mechanism.  Heat losses due to radiation effectively leads to an additional flow of heat through the sample, and makes the thermal conductivity appear too large.  Appropriately choosing the aspect ratio of sample can help to reduce this error \cite{HeikesUre}.

The final assumption (iii) that the flows of heat and electricity may be treated as being effectively one-dimensional, does not seem to have been considered previously.  The one dimensionality relies the contacts being made in such a way that the temperature and voltage are uniform across the contact faces of the sample.  In Harman's original work it was noted that this could be achieved by applying a thin layer of solder over the contact faces of the sample.  The high electrical and thermal conductivity of the solder ensures assumption (iii) is valid.  However, one could ask what would happen if this condition were not satisfied, either through error (\textit{i.e.} the solder does not completely cover the surface of the sample, or does not make a good contact in some places), or by deliberately using small contacts.  It is reasonable to assume that this will lead to more complicated distributions of temperature and voltage in the sample.  If the thermopower is anisotropic, we have already seen that new effects can occur when the temperature and voltage distributions become non-trivial, both in the form of eddy currents and the Bridgman effect.  Even if the thermopower is isotropic, we will see that electrical and thermal anisotropy can also have interesting consequences.  A further degree of freedom is the geometry of the sample itself, and the particular arrangement of current and voltage contacts on its surface.  In the next section, we introduce the model which will be used to investigate the effects of small contacts and anisotropy in the Harman method.

\begin{figure}[h]
\begin{center}
\includegraphics[width=0.4\textwidth]{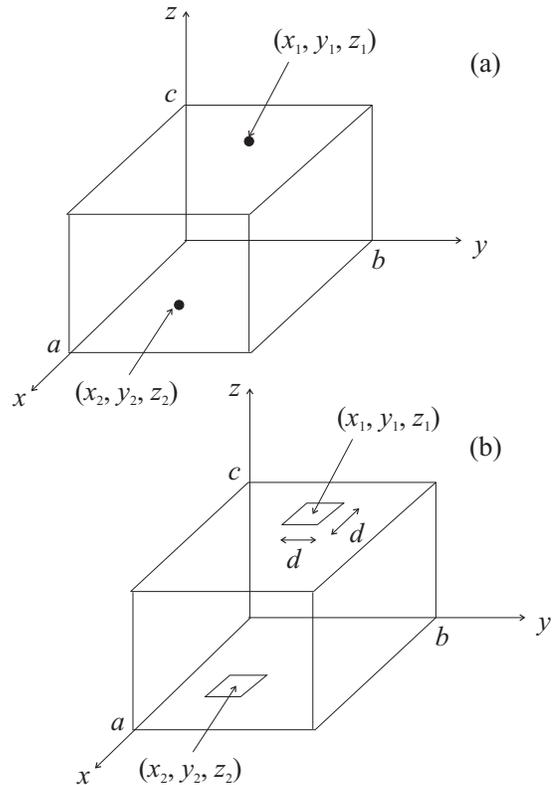}
\caption{\label{fig:3dsample} Schematic representation of the sample geometry.  All faces are electrically and thermally insulated.  An electrical current $I$ is passes into the sample through the contact at $(x_1,y_1,z_1)$, and leaves the sample through an identical contact at $(x_2,y_2,z_2)$. (a) Point current contacts; (b) Current contacts with a finite area $d^2$.} 
\end{center}
\end{figure}

\section{Model}\label{sec:model}

We seek a minimal model which is capable of exploring the effects of small current contacts and transport coefficient anisotropy on the temperature and voltage distributions created in the Harman method.  We begin with a description of the sample configuration.  The sample is chosen to be a cuboid with dimensions $a \times b \times c$ in the $(x,y,z)$ directions respectively.  We choose a coordinate system which coincides with one corner of the sample, as illustrated in figure \ref{fig:3dsample}.  The sample material is taken to be homogeneous, but may have arbitrary degrees of anisotropy in all of its transport coefficients. 

The sample is connected to two current leads, which have a thermopower $S_l$, and which we assume to have negligible thermal conductance.  Hence, there is no heat loss from the sample along the current leads.  The leads connect to the sample through square contacts of dimensions $d \times d$.  Each contact may be placed on any face of the sample.  The boundary conditions at the current contacts need careful consideration.  In reality, the contact is likely to be made from a blob of solder, which will have a very high electrical and thermal conductivity.  The region under the contact will therefore be approximately equipotential and isothermal.  The electrical and heat current density will vary under the contact to ensure that these conditions are satisfied.  Away from the current contacts, all surfaces of the sample are taken to be electrically and thermally insulated.  The disadvantage of these boundary conditions is that they make the problem very difficult to handle analytically.  The reason is that the boundary conditions on the contact faces are not uniform, since over one part of the surface the function itself is specified, while over the rest the derivative is specified.  To avoid this complication, we adopt a different boundary condition, namely that the normal components of the electrical and heat current density under the contacts are uniform.  This means that the problem has Neumann boundary conditions on all surfaces, and can be solved by separation of variables.  In the limit of point contacts, the two choices of boundary condition are in some sense equivalent, although the voltage and temperature are infinite at the point contacts.  For small contacts (\textit{i.e.} much smaller than the cross-sectional area of the sample), the choice of boundary conditions is not expected cause large changes in the solution in parts of the sample away from the contact.  Similarly for large contacts (approximately the same area as the sample) the variation of the temperature and voltage in the contact region is expected to be small.  Finally, we will show later although the variation of the temperature and voltage under the contacts is unphysical, the average temperature and voltage underneath the contacts does have a physical significance.

Since we require a minimal model only, we will not attempt to solve the formidable non-linear problem posed by equations (\ref{eq:eleccont}) and (\ref{eq:heatcont}).  Instead, take the following approach to linearise these equations.  We first write the temperature and electrical current density in terms of symmetric and anti-symmetric parts with respect to the injected current $I$; $\mathbf j(I) = \mathbf j^s(I) + \mathbf j^a(I)$, $T(I) = T^s(I) + T^a(I)$, with $\mathbf j^s(-I) = \mathbf j^s(I)$, $\mathbf j^a(-I) = -\mathbf j^a(I)$ \textit{etc}.  Now consider expanding $\mathbf j$ and $T$ in powers of $I$.  Clearly, we have $\mathbf j^a(I) = \mathcal O(I)$, $\mathbf j^s(I) = \mathcal O(I^2)$, $T^a(I) = \mathcal O(I)$ and $T^s(I) = \bar T + \mathcal O(I^2)$.  Substituting these expressions into equations (\ref{eq:eleccont}) and (\ref{eq:heatcont}), and retaining only terms of $\mathcal O(I)$, we find
\begin{eqnarray}
\label{eq:linelec}
&\displaystyle{\parone{j^a_i}{x_i} = 0}, \\
\label{eq:linheat}
&\displaystyle{-\parone{}{x_i}\Big ( \kappa_{ik}\parone{T^a}{x_k}\Big) = -\bar T S_{ik}\parone{j^a_k}{x_i}},
\end{eqnarray}    
with
\begin{equation}
j^a_i = \sigma_{ik}\parone{\chi^a}{x_k} - \sigma_{il}S_{lk}\parone{T^a}{x_k}.
\end{equation}
All transport coefficients are to be evaluated at the ambient temperature $\bar T$.  Note that the internal Peltier effect is absent, since we have restricted attention to homogeneous samples.  The Thomson effect and the Joule heating also do not appear, since they are both $\mathcal O(I^2)$ effects. Further, they would only alter the anti-symmetric parts of $\chi$ and $T$ at $\mathcal O(I^3)$.  Hence, by working with $T^a$ and $\chi^a$, our approximation holds to $\mathcal O(I^3)$.  In the remainder of this paper we will drop the `$a$' superscript, but it should be remembered that when comparing with experiments the anti-symmetric parts of $\chi$ and $T$ should be taken.  As discussed in section \ref{sec:Harman}, this procedure is usually adopted in experiments anyway in order to remove extraneous effects of Joule heating associated with contact resistance or misplacement.

\section{Solution}\label{sec:gensoc} 

\subsection{General Solution and the Reciprocity Theorem}\label{sec:flow2} 

We will now find the most general solution to equations (\ref{eq:eleccont}) and (\ref{eq:heatcont}).  In calculating the voltage distribution in an isotropic sample with point electrical contacts on the corners, Logan \textit{et al} \cite{Logan} used an analogy with electrostatics to obtain the solution.  Each point contact acts as a source of current, much like a point charge acts as a source of electrical flux.  An infinite number of image charges must be introduced to enforce the boundary conditions on the surfaces of the sample.  The voltage distribution in the sample is then given by the sum of the potentials due to each image charge.

We will adopt a similar method to solve equations (\ref{eq:linelec}) and (\ref{eq:linheat}).  To illustrate the method, we consider two point current contacts located at positions $(x_1,y_1,z_1)$ and $(x_2,y_2,z_2)$.  In practice, these points must lie on the surface of the sample, but for argument sake we consider them to lie in the interior of the sample.  One contact is a source of current, while the other is a sink.  Defining $\Delta(x,y,z) = \delta(x-x_1)\delta(y-y_1)\delta(z-z_1) - \delta(x-x_2)\delta(y-y_2)\delta(z-z_2)$, equations (\ref{eq:linelec}) and (\ref{eq:linheat}), now written-out in full, are modified to read
\begin{widetext}
\begin{equation}\label{eq:eleccont2}        
  \sigma_x \partwo{\chi}{x} + \sigma_y \partwo{\chi}{y} + \sigma_z \partwo{\chi}z  - \sigma_x S_x \partwo{T}{x} - \sigma_{y} S_{y} \partwo{T}{y} - \sigma_z S_z \partwo{T}{z} = I\Delta(x,y,z),
\end{equation}
\begin{multline}\label{eq:heatcont2}  
 -\big (\kappa_x+ \bar T\sigma_{x} S_{x}^2 \big)\partwo{T}{x} - \big (\kappa_y+ \bar T\sigma_{y} S_{y}^2 \big)\partwo{T}{y} - \big (\kappa_z+ \bar T\sigma_{z} S_{z}^2 \big)\partwo{T}{z} + \bar T \Big ( \sigma_x S_x \partwo{\chi}{x} + \sigma_{y} S_{y} \partwo{\chi}{y} +\sigma_{z} S_{z} \partwo{\chi}{z} \Big )  = \Pi_lI\Delta(x,y,z).
\end{multline}
\end{widetext}
The right-hand side of equation (\ref{eq:eleccont2}) represents a point source of current at $(x_1,y_1,z_1)$, and a point sink of current at $(x_2,y_2,z_2)$.  The right-hand side of equation (\ref{eq:heatcont2}) represents a point source of Peltier heat at $(x_1,y_1,z_1)$, and a point sink of Peltier heat at $(x_2,y_2,z_2)$. Rather than attempt to explicitly find the locations of the image charges, we adopt a different approach.  We expand $\chi$ and $T$ using a basis of functions whose derivatives vanish at the sample boundaries:
\begin{align} 
\chi &= \sum_{mnp}a_{mnp}\coss{\tilde m}{x}\coss{\tilde n}{y}\coss{\tilde p}{z}, \nonumber \\
T &= \sum_{mnp}b_{mnp}\coss{\tilde m}{x}\coss{\tilde n}{y}\coss{\tilde p}{z}, \nonumber 
\end{align}
with $\tilde m = m\pi/a$, $\tilde n= n \pi/b$ and $\tilde p= p \pi/c$, and where the sums run over all integers, $-\infty < m,n,p < \infty$.  It is clear from equations (\ref{eq:anisoelec}) and (\ref{eq:anisoheat}) that by requiring that the derivatives of $\chi$ and $T$ vanish independently on all boundaries, the sample is electrically and thermally insulated.  Similarly, $\Delta(x,y,z)$ may also be written in this way, using the following representation of the delta function \cite{comment2}:
\begin{align}
\delta(x - x_1) = \frac{1}{a}\sum_{m} \coss{\tilde m}{x_1} \coss{\tilde m}{x}.
\end{align}
We now substitute these representations into equations (\ref{eq:eleccont2}) and (\ref{eq:heatcont2}), which reduce to two simultaneous algebraic equations for $a_{mnp}$ and $b_{mnp}$.  Omitting the algebraic details, we find for $\chi$ and $T$

\begin{widetext}
\begin{align}
\label{eq:bigchi}
& \chi(x,y,z) = I \sum_{mnp}'\frac{\coss{\tilde m}{x_2}\coss{\tilde n}{y_2}\coss{\tilde p}{z_2}-\coss{\tilde m}{x_1}\coss{\tilde n}{y_1}\coss{\tilde p}{z_1}}{abc}\coss{\tilde m}{x}\coss{\tilde n}{y}\coss{\tilde p}{z} \times \nonumber \\
&\frac{\big(\kappa_x - \sigma_xS_x(\Pi_l-\Pi_x)\big)\tilde m^2 + \big(\kappa_y - \sigma_yS_y(\Pi_l-\Pi_y)\big)\tilde n^2 + \big(\kappa_z - \sigma_zS_z(\Pi_l-\Pi_z)\big)\tilde p^2}{(\sigma_x \tilde m^2 + \sigma_y \tilde n^2 + \sigma_z\tilde p^2)(\kappa_x \tilde m^2 + \kappa_y \tilde n^2 + \kappa_z \tilde p^2) + \bar T \big( \sigma_x\sigma_y(S_x-S_y)^2 \tilde m^2 \tilde n^2 + \sigma_x\sigma_z(S_x-S_z)^2 \tilde m^2 \tilde p^2 + \sigma_y\sigma_z(S_y-S_z)^2 \tilde n^2 \tilde p^2\big)}, \\
\label{eq:bigT}
& T(x,y,z) = -I \sum_{mnp}'\frac{\coss{\tilde m}{x_2}\coss{\tilde n}{y_2}\coss{\tilde p}{z_2}-\coss{\tilde m}{x_1}\coss{\tilde n}{y_1}\coss{\tilde p}{z_1}}{abc}\coss{\tilde m}{x}\coss{\tilde n}{y}\coss{\tilde p}{z} \times \nonumber \\
&\frac{\sigma_x(\Pi_l - \Pi_x)\tilde m^2 + \sigma_y(\Pi_l - \Pi_y)\tilde n^2 + \sigma_z(\Pi_l - \Pi_z)\tilde p^2}{(\sigma_x \tilde m^2 + \sigma_y \tilde n^2 + \sigma_z\tilde p^2)(\kappa_x \tilde m^2 + \kappa_y \tilde n^2 + \kappa_z \tilde p^2) + \bar T \big( \sigma_x\sigma_y(S_x-S_y)^2 \tilde m^2 \tilde n^2 + \sigma_x\sigma_z(S_x-S_z)^2 \tilde m^2 \tilde p^2 + \sigma_y\sigma_z(S_y-S_z)^2 \tilde n^2 \tilde p^2\big)}.
\end{align}
\end{widetext}
The prime above the sum indicates that the term with $m=n=p=0$ is not included.  Physically, this is because there is no net current/heat being delivered to the sample.  Although these expressions may seem somewhat cumbersome, the elegance of this representation is that it emphasises the symmetry of the solutions with respect to the three coordinate directions.  Only the choice of the current contact positions breaks this symmetry.

Another advantage of this representation is that it enables us to easily investigate a property of the solution known as \textit{reciprocity}.  Define the quantity $V_{ABCD}$ as the potential difference measured between points $C = (x,y,z)$ and $D=(x',y',z')$, with a current $I_{AB}$ entering the sample at $A$ and leaving at $B$.  The reciprocity theorem states that the voltage is symmetric under interchange of voltage and current contacts, or $V_{ABCD} = V_{CDAB}$, provided that total current passed is the same, \textit{i.e.} $I_{AB} = I_{CD}$.  This theorem is known to hold for the ac voltage \textit{i.e.} in the absence of thermoelectric effects.  The form of equation (\ref{eq:bigchi}) indicates that the theorem continues to hold in the presence of (anisotropic) thermoelectric effects.  Further, it is clear from equation (\ref{eq:bigT}) that this reciprocity also holds for the temperature difference, $\Delta T_{ABCD} = \Delta T_{CDAB}$.  That reciprocity continues to hold in this problem is not surprising, given that the linear response condition required for its existence is satisfied here \cite{Casimir}.

Once the particular faces on which the contacts are located have been chosen, it is possible to simplify the form of solution by performing one of sums appearing in (\ref{eq:bigchi}) and (\ref{eq:bigT}).  There are three distinct contact configurations on a cuboid.  These have the contacts on the same face, on opposite faces, and on adjacent faces of the sample.  In this work, we will primarily be interested in the first two of these configurations.  The choice of particular faces for the calculations is of course arbitrary; here we will choose the faces to to those at $z=0,c$. Although the solution with contacts on adjacent faces follows naturally from equations (\ref{eq:bigchi}) and (\ref{eq:bigT}) with appropriate choices of contact location, it is not easy to simplify the solution into a compact form.  Hence, we will not explicitly consider this configuration in this paper.  The solutions for contacts on the same side and on opposite sides of the sample may be found in appendices \ref{sec:same} and \ref{sec:opp}.  In the next section, we will explore the general form of the solution by considering the voltage distribution due to an alternating current.  

\subsection{The ac Voltage}

A quantity of particular interest for each contact configuration is the voltage distribution in the sample due to an alternating current.  As discussed in section \ref{sec:Harman}, the Peltier effects at the current contacts are unable to change the temperature distribution in the sample if the current oscillates with sufficiently high frequency.  Since the temperature in the sample remains uniform (at leading order in $I$), all thermoelectric effects vanish.  The ac voltage distributions for each contact configuration are obtained from equations (\ref{eq:sameside}) and (\ref{eq:oppside}) by setting all thermoelectric coefficients equal to zero:
\begin{equation}
\label{eq:acoppside}
\displaystyle{\chi^o_{\mathrm{ac}} = \frac{Ic}{\sigma_z ab}\Big(\frac{z}{c}-\frac{1}{2} \Big ) + \frac{I}{\sigma_z d^2} \sum_{mn}'\Omega_{mn}(x,y,z;R_{mn})},
\end{equation}
\begin{equation}\label{eq:acsameside}
\displaystyle{\chi^s_{\mathrm{ac}} = \frac{I}{\sigma_z d^2} \sum_{mn}'\Lambda_{mn}(x,y,z;R_{mn})},
\end{equation}
where $R_{mn}^2 = r_{xz}^2\tm^2 + r_{yz}^2\tn^2$.  Here, $\chi^s_{\mathrm{ac}}$ is the ac voltage distribution in a sample whose current contacts are on the same face at $z=0$, while $\chi^o_{\mathrm{ac}}$ is the ac voltage distribution in a sample whose current contacts are on opposite faces at $z=0$ and $z=c$.  The current contacts each have a finite area $d^2$.  The solution for $\chi^o_{\mathrm{ac}}$ divides into two parts.  The first term corresponds to a uniform flow of electrical current through the sample.  The sum represents the modification to this uniform flow due to the small current contacts.  The ac voltage has a minimum over the contact where the current enters the sample, and a maximum over the contact where the current leaves.  The maximum (minimum) values of the voltage occur at the centers of the current contacts, and increase (decrease) as the contact size $d$ tends to zero.  The scale over which the voltage distribution is modified in the $z$-direction is controlled by the electrical anisotropies $r_{xy}$ and $r_{yz}$.  Larger anisotropy leads to the voltage distribution being modified only very close to the contact faces.    

Very similar statements can be made for $\chi^s_{\mathrm{ac}}$.  The main difference from the previous case is that there is no linear term in the solution.  This is simply because there is no uniform flow of electricity in the $z$-direction; the current enters through one contact on the face at $z=0$, and leaves through another contact also at $z=0$.

\section{Harman Figure of Merit}\label{sec:hfom}

In section \ref{sec:Harman}, we showed by direct calculation that the Harman FOM given by equation (\ref{eq:figofmerit}) is equal to $\bar T (S_l - S_z)^2\sigma_z/\kappa_z$, which is approximately equal to the intrinsic FOM in the $z$-direction, $z_z$.  Here, we will simply \textit{define} the Harman FOM by the equation $z_H = (V_{\mathrm{dc}} - V_{\mathrm{ac}})/V_{\mathrm{ac}}$, and compute this quantity using the results of the previous section.  

It is obvious that once the temperature and voltage distributions vary with position on the contact faces, the Harman FOM is no-longer a unique quantity.  In general, it is a function of twelve coordinates; the two current contact locations, and the two voltage contact locations.  It is also generally true that the magnitude of the Harman FOM will be equal to zero for certain configurations, and tend to infinity for others.  This simply because for given contact contact locations it will generally be possible to find voltage contact locations for which either $V_{\mathrm{dc}} - V_{\mathrm{ac}}=0$ with $V_{\mathrm{ac}} \ne 0$, or  $V_{\mathrm{ac}} = 0$ with $V_{\mathrm{dc}} - V_{\mathrm{ac}} \ne 0$. 

Rather than state the results in full generality, it is instructive to start with the simplest case first. 
 
\subsection{Large Contacts}

We start with the case where the sample has a square cross-section $(a=b)$, and the contact size is made equal to the area of the sample, $d=a$.  Clearly, this case is only applicable when the contacts are on opposite sides of the sample.  Setting $d=b=a$, we find that all of the Fourier coefficients (\ref{eq:fourierco}) apart from $\nu_{00}$ and $\gamma_{00}$ vanish identically.  This means that the sums appearing in equations (\ref{eq:oppside}) are also identically zero. Only the linear terms remain, the expressions for $\chi^o$ and $T^o$ being identical to (\ref{eq:1dchi}) and (\ref{eq:1dtemp}) respectively, up to the constant mean temperature and voltage which we have neglected.  Hence the temperature and voltage are both uniform over the contact faces.  Following through the analysis presented in section \ref{sec:Harman}, the Harman FOM reduces to $z_H = \bar T (S_l-S_z)^2\sigma_z/\kappa_z$, which is the expected value.  Note that this result does not require any assumptions about the anisotropy; large current contacts are sufficient to recover the normal Harman FOM.

\subsection{Isotropic $S$, Anisotropic $\sigma$ and $\kappa$}

The next case is when the thermopower is taken to be isotropic in all directions, $S_x=S_y=S_z=S$, while the electrical and thermal conductivity are still anisotropic.  Although this situation is unlikely to occur exactly in practice, it is theoretically possible.  The simplification when the thermopower is isotropic is that the Bridgman effect does not occur.  Hence, the right-hand side of equation (\ref{eq:linheat}) is zero, and the temperature distribution becomes completely de-coupled from the voltage.   From equations (\ref{eq:sameside}) and (\ref{eq:oppside}), the temperature distributions in the sample are given by
\begin{multline}\label{eq:tempoppside1}
T^o =  - \frac{(\Pi_l - \Pi) Ic}{\kappa_z ab}\Big(\frac{z}{c}-\frac{1}{2} \Big )  \\
- \frac{(\Pi_l - \Pi) I}{\kappa_z d^2} \sum_{mn}'\Omega_{mn}(x,y,z;K_{mn}),
\end{multline}
\begin{align}\label{eq:tempsameside1}
T^s =  - \frac{(\Pi_l - \Pi) I}{\kappa_z d^2} \sum_{mn}'\Lambda_{mn}(x,y,z;K_{mn}).
\end{align}
with $K_{mn}^2 = k_{xz}^2 \tm^2 + k_{yz}^2 \tn^2$.  The forms of these expressions are identical to the corresponding equations for the ac voltage, (\ref{eq:acsameside}) and (\ref{eq:acoppside}).  The key difference is that the scale over which the non-uniform part of the temperature distribution varies in the $z$-direction is controlled by the thermal anisotropies $k_{xz}$ and $k_{yz}$, rather than the electrical anisotropies.  Hence although the mathematical forms of the ac voltage and temperature distributions may be qualitatively similar, electricity and (thermally conducted) heat can flow in very different ways if the electrical and thermal anisotropies are different.  

\begin{figure}[h!]  
\subfigure[\ $\mathbf j$]{\includegraphics[width=0.5\textwidth]{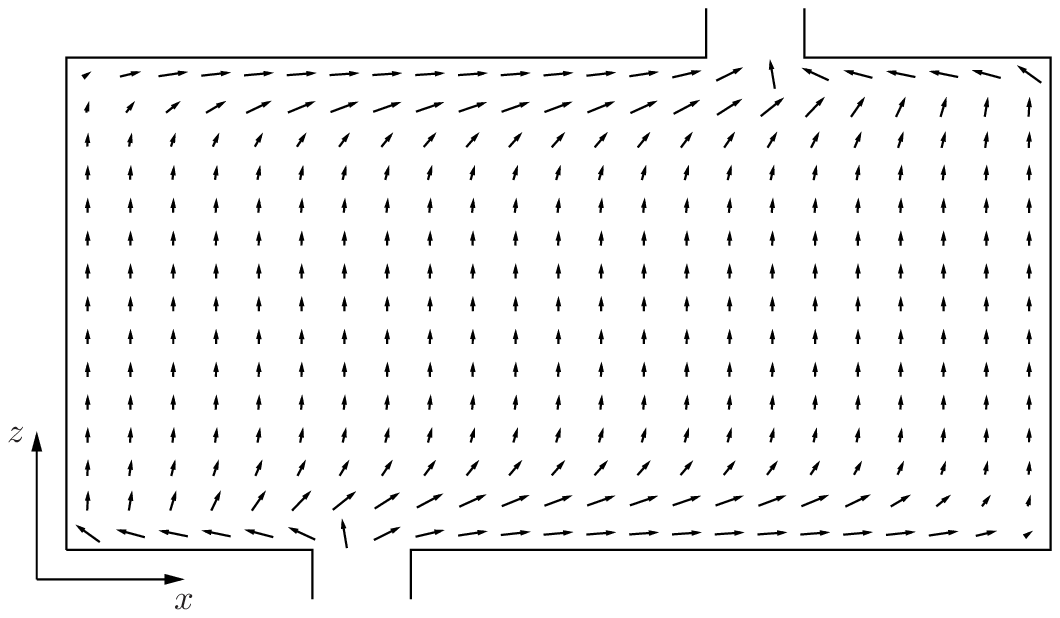}}
\subfigure[\ $\mathbf j_Q - \Pi \mathbf j$]{\includegraphics[width=0.5\textwidth]{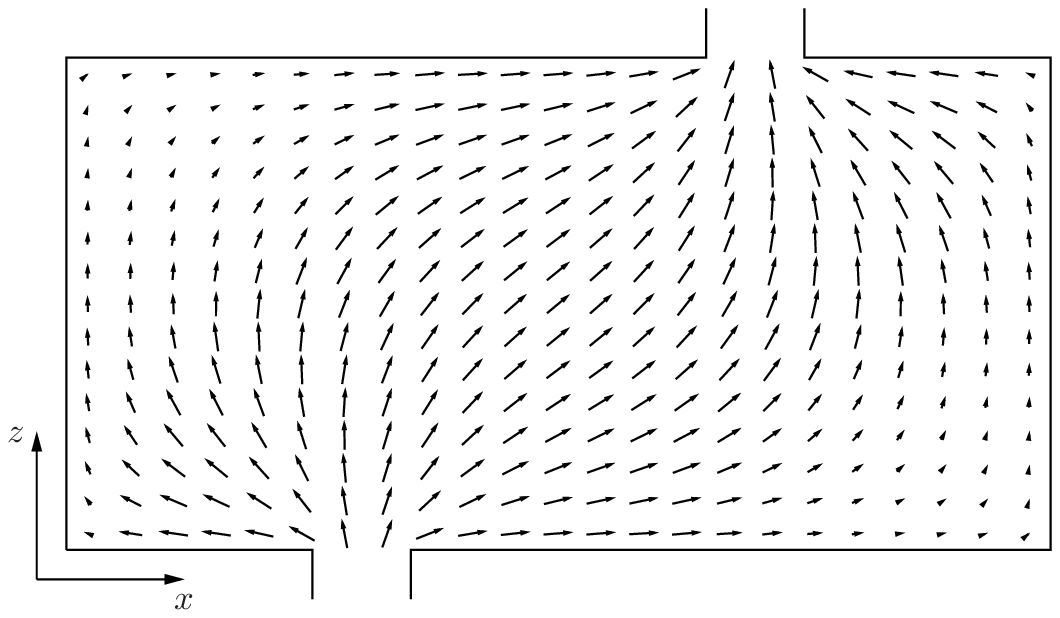}}     
\caption{Plots showing the flow patterns of the electrical and heat currents in a cross-section at $y=0.5$ through a particular sample.  The sample has dimensions $a=1$, $b=1$, $c=0.5$, with current contacts at $(0.5a,0.5b,0)$ and $(0.5a,0.5b,c)$.  The electrical and thermal anisotropies are $r=10$, $k=1$.  The arrows indicate the local magnitude and direction of the current density, with the magnitude capped at $2I/ab$ ($2(\Pi-\Pi_l)I/ab$ for the heat current) for clarity.  (a) Electrical current $\mathbf j$; (b) Heat current due to thermal conduction, $\mathbf j_Q - \Pi\mathbf j$;  }
\label{fig:flows}
\end{figure}

To illustrate this, figure \ref{fig:flows}(a) shows the flow pattern for the electrical current in a section through a simple sample geometry.  Figure \ref{fig:flows}(b) shows the part of the heat current due to thermal conduction, $\mathbf j_Q - \Pi\mathbf j$.   The sample is thermally isotropic ($k_{xz}=k_{yz}=k=1$), while the electrical anisotropy is $r_{xz}=r_{yz}=r=10$.  While the heat flow is confined to a region in the middle of the sample, the large electrical anisotropy causes the electrical current to spread-out rapidly, leading to an almost uniform flow in the bulk.  The non-uniform part of the flow occurs only in regions with a width of order $a/r$ next to the contact faces.  Note that the size of the contact and aspect ratio $c/a$ of the sample are also important here.  Smaller contacts and aspect ratio make the flows more confined, while larger contacts and aspect ratio leads to uniform flow over a greater part of the sample.  

Isotropic thermopower also guarantees the absence of thermoelectric eddy currents in the sample.  This means that the dc voltage and temperature distributions in the sample are related by the usual expression,  $\chi = \chi_{\mathrm{ac}} + ST$. 

Let us explicitly construct the Harman FOM in this case.  From equation (\ref{eq:acsameside}) or (\ref{eq:acoppside}) (depending on the current contact locations),  the ac potential difference measured between two points $(x,y,z)$ and $(x',y',z')$ is $V_{\mathrm{ac}} = \chi_{\mathrm {ac}}(x',y',z') - \chi_{\mathrm {ac}}(x,y,z) = IR^{\mathrm m}_{z}$, where $R^{\mathrm m}_{z}$ is the measured electrical resistance to alternating current between the voltage contacts.  From equation (\ref{eq:tempsameside1}) or (\ref{eq:tempoppside1}) the temperature difference between the same two points is just $\Delta T = T(x',y',z') - T(x,y,z) = -(\Pi_l-\Pi)I/K^{\mathrm m}_{z}$, where $K^{\mathrm m}_{z}$ is the measured thermal conductance between the voltage contacts.  Since the relation $\chi = \chi_{\mathrm{ac}} + ST$ holds in this case, the measured dc potential difference (including the contribution of the leads) is given by $V_{\mathrm{dc}} = V_{\mathrm{ac}}  - (S_l - S)\Delta T$.  Hence, the Harman FOM may be written as 
\begin{equation}\label{eq:isofom}
z_H^s = \frac{\bar T (S_l - S)^2}{R^{\mathrm m}_{z} K^{\mathrm m}_{z}}.
\end{equation}
As expected, the Harman FOM is not a unique number, and depends on the voltage contact locations through $R^{\mathrm m}_{z}$ and $K^{\mathrm m}_{z}$.  However, the physical meaning of the Harman FOM survives in this case, since it can be related to a measured electrical resistance and thermal conductance.

Finally, if the electrical and thermal anisotropy in each direction are identical (\textit{i.e.} $r_{xz} = k_{xz}$ and $r_{yz} = k_{yz}$), then $R_{mn} = K_{mn}$.  Such a situation might occur in a metal where the Weidemann-Franz law holds, and the electronic contribution to the thermal conductivity dominates.  It is clear from equations (\ref{eq:acoppside}), (\ref{eq:acsameside}), (\ref{eq:tempoppside1}) and (\ref{eq:tempsameside1}) that the Harman FOM reduces to $z_H = \bar T (S_l-S_z)^2\sigma_z/\kappa_z$, independent of the details of the current contact configuration.  Essentially, the effects of the small current contact modify the electrical resistance and thermal conductance in the same way, resulting in no net change to the FOM.     

\subsection{Anisotropic Thermopower and Tetragonal Symmetry; $x \equiv y \not\equiv z$}\label{sec:q2d1}

We will now consider how anisotropic thermopower changes the conclusions of the previous section.  It is convenient to restrict attention to a case where the crystal structure of the material has a tetragonal symmetry with $x \equiv y \not\equiv z$.  This provides a minimal level of anisotropy which still captures the effects we wish to describe here.  It will also form a useful basis for the discussion of the experiments of Tamura \textit{et al} in section \ref{sec:case}.  Tetragonal symmetry means that the sample is isotropic in two directions, but anisotropic in the third.  We consider here the case where the isotropic directions lie parallel to the contact faces.  In our geometry, the transport coefficients in the $x$ and $y$-directions are assumed to be the same, but take different values to those in the $z$-direction.  Hence, the anisotropy parameters satisfy $r_{xz}=r_{yx}=r$, $s_{xz} = s_{yz} = s$, $k_{xz} = k_{yz} = k$.  We can extract the $m$ and $n$ dependence from the quantities $u_{mn}$, $v_{mn}$, $w_{mn}$ and $\lambda_{i}$, each of which is now proportional to $\omega_{mn} = \sqrt{\tm^2 + \tn^2}$.  It is also convenient to define the following combinations of parameters
\begin{equation}
\begin{array}{cc}
\alpha = z_z^{-1}k^2 - r^2s(s_l-s),  &\beta = s_l-1-z_z^{-1}, \\ 
\epsilon = -r^2(s_l-s), &\delta = s_l -1, \\
\alpha' = \alpha - s_l \epsilon, & \beta' = \beta - s_l \delta. 
\end{array}
\end{equation}
The solutions for each contact configuration are given by
\begin{widetext}
\begin{align}\label{eq:q2dsameside}
\left(
\begin{array}{c}
\chi^s \\
S_z T^s
\end{array}
\right) &=
\frac{I z_z}{\sigma_z d^2} \sum_{i=1,2}\frac{(-1)^{i+1}}{\mu_2^2-\mu_1^2}
\left(
\begin{array}{c}
\alpha + \beta \mu_i^2 \\
\epsilon + \delta \mu_i^2
\end{array}
\right)
\sum_{mn}'
\Lambda_{mn}(x,y,z;\lambda_i),
\end{align}
\begin{equation}\label{eq:q2doppside}
\left(
\begin{array}{c}
\chi^o \\
S_z T^o
\end{array}
\right) =
\frac{Ic}{\sigma_zab}\Big(\frac{z}{c}-\frac{1}{2}\Big)
\left(
\begin{array}{c}
1-\sigma_zS_z(\Pi_l-\Pi_z)/\kappa_z \\
-\sigma_zS_z(\Pi_l-\Pi_z)/\kappa_z
\end{array}
\right) + 
\frac{Iz_z}{\sigma_z d^2} \sum_{i=1,2}\frac{(-1)^{i+1}}{\mu_2^2-\mu_1^2}
\left(
\begin{array}{c}
\alpha + \beta \mu_i^2 \\
\epsilon + \delta \mu_i^2
\end{array}
\right)
\sum_{mn}' \Omega_{mn}(x,y,z;\lambda_i),
\end{equation}
\begin{equation}
\label{eq:mu1}
\lambda_i^2 = \omega_{mn}^2\mu_i^2 = \frac{\omega_{mn}^2}{2}\bigg( r^2+k^2+r^2z_z(1-s)^2 +(-1)^{i+1}\sqrt{\big(r^2+k^2+r^2z_z(1-s)^2 \big)^2-4r^2k^2} \bigg ).
\end{equation}
\end{widetext}
We note the following important features of this solution. First, the voltage and temperature distributions both depend on all of the transport coefficients through $\mu_i$.  Previously, the temperature distribution only depended on the thermal anisotropy through $K_{mn}$. Second, the voltage distributions cannot be written in the form $\chi = \chi_{\mathrm{ac}} + ST$.  These features occur as a direct result of the Bridgman effect and thermoelectric eddy currents, which couple the flows of heat and electricity together in a non-trivial way.   

We will now construct the Harman FOM for current contacts on the same side of the sample.  The ac potential difference again follows from equation (\ref{eq:acsameside}) with $R_{mn} = r\omega_{mn}$, and defines a measured resistance to alternating current between the voltage contacts.  The temperature difference between the voltage contacts is obtained from equation (\ref{eq:q2dsameside}):
\begin{equation}
\Delta T =  \frac{I z_z}{\sigma_z S_z d^2} \sum_{i=1,2}\frac{(-1)^{i+1}}{{\mu_2^2 - \mu_1^2}}(\epsilon + \delta \mu_i^2)\sum_{mn}'\Delta \Lambda_{mn}(\mu_i\omega_{mn}).
\end{equation}
It is at this point that we begin to see the complications introduced by the anisotropic thermopower.  Unlike before, we cannot formally extract a measured thermal conductance from this expression.  The reason is that the Peltier coefficient is not a unique quantity for the sample when the thermopower is anisotropic.  One might be tempted to write this expression as $\Delta T= \Pi^{\mathrm m}_{z} I/K^{\mathrm m}_{z}$, which defines both a measured thermal conductance and Peltier coefficient.  However, there is little value in this approach, since the measured dc potential difference, given by  
\begin{equation}
V_{\mathrm{dc}} =  \frac{I z_z}{\sigma_z d^2} \sum_{i=1,2}\frac{(-1)^{i+1}}{{\mu_2^2 - \mu_1^2}}(\alpha' + \beta' \mu_i^2)\sum_{mn}'\Delta \Lambda_{mn}(\mu_i\omega_{mn})
\end{equation}
cannot be trivially related to either of these quantities.  Hence the most we can say is that the measured Harman FOM is given by equation (\ref{eq:figofmerit}):
\begin{equation}\label{eq:q2dsamefom}
z_H^s = z_z \frac{\displaystyle{\sum_{i=1,2}\frac{(-1)^{i+1}}{{\mu_2^2 - \mu_1^2}}(\alpha' + \beta'\mu_i^2)\sum_{mn}'\Delta \Lambda_{mn}(\mu_i\omega_{mn})}}{\displaystyle{\sum_{mn}'\Delta \Lambda_{mn}(r\omega_{mn})}}-1.
\end{equation}
Similarly, for current contacts on opposite faces of the sample we have 
\begin{widetext}
\begin{equation}\label{eq:q2doppfom}
z_H^o = \frac{\displaystyle{\frac{\bar T(S_l - S_z)^2\sigma_z}{\kappa_z}\Big (\frac{z'-z}{c}\Big) + 
\frac{ab}{cd^2} \Bigg (z_z\sum_{i=1,2}\frac{(-1)^{i+1}}{{\mu_2^2 - \mu_1^2}}(\alpha' + \beta'\mu_i^2)\sum_{mn}'\Delta \Omega_{mn}(\mu_i \omega_{mn}) - \sum_{mn}'\Delta \Omega_{mn}(r\omega_{mn}) \Bigg )}}{\displaystyle{\Big (\frac{z'-z}{c} \Big ) + \frac{ab}{cd^2}\sum_{mn}'\Delta\Omega_{mn}(r\omega_{mn})}},
\end{equation}
\end{widetext}
which again cannot be related in any simple way to a measured electrical resistance, thermal conductance and Peltier coefficient.

We conclude that when the thermopower is anisotropic, the Harman FOM becomes even more complicated.  As well as the dependence on contact locations, we cannot even interpret it in terms of `measured' parameters, as we could when the thermopower was isotropic. The principle reason for this is the breakdown of the relation $\chi = \chi_{\mathrm{ac}} + ST$, which makes it impossible to separate the thermoelectric part of the voltage from the Ohmic part. In the general case, it would be almost impossible to extract any useful information from the Harman FOM.  However, the degrees of freedom associated with the contact locations and the sample geometry can be used to simplify the formulae.  We will consider some of these limits in the next section. 

\subsection{Limit of a Very Thick or Thin Sample}\label{sec:thickthin} 

The solution does take on particularly simple forms in certain limits, namely those of a thick sample, and a thin sample. For simplicity, we consider a sample with a square cross-section, and hence $a=b$.  We take as an example the case where both voltage contacts are placed on the same face as the current contacts, $z=z'=0$.  The FOM for this configuration is given by    
\begin{widetext}
\begin{equation}
z_H^s = z_z \frac{\displaystyle{\sum_{i=1,2}\frac{(-1)^{i+1}}{{\mu_2^2 - \mu_1^2}}(\alpha' + \beta'\mu_i^2)\sum_{mn}' (\gamma_{mn} - \nu_{mn})(\coss{\tm}{x'}\coss{\tn}{y'} -\coss{\tm}{x}\coss{\tn}{y})\frac{\coth \mu_i \omega_{mn}c }{\mu_i \omega_{mn}}}}{\displaystyle{\sum_{mn}'(\gamma_{mn} - \nu_{mn})(\coss{\tm}{x'}\coss{\tn}{y'} -\coss{\tm}{x}\coss{\tn}{y})\frac{\coth r\omega_{mn}c}{r\omega_{mn} } }}-1.
\end{equation}
\end{widetext}
Now consider the arguments of the hyperbolic functions.  For $c/a \rightarrow \infty$, the argument is large for all $m$ and $n$, and hence we can approximate $\coth \mu_i \omega_{mn} c/a \sim 1$, $\coth r\omega_{mn} c/a \sim 1$ for all $m$, $n$.  In the opposite limit $c/a \rightarrow 0$, terms with small $m$ and $n$ can be approximated using $\coth \mu_i\omega_{mn} c/a \sim a/(\mu_i \omega_{mn}r)$, $\coth r\omega_{mn} c/a \sim a/(r \omega_{mn}c)$.  This approximation will fail for large enough $m$ and $n$, but such terms are irrelevant since the sum is always dominated by terms with small $m$ and $n$.

Taking the limit $c/a \rightarrow \infty$, we find
\begin{multline}\label{eq:samesamefom}
z_H^s \rightarrow rz_z \sum_{i=1,2}\frac{(-1)^{i+1}}{{\mu_2^2 - \mu_1^2}}\Big (\frac{\alpha'}{\mu_i} + \beta'\mu_i \Big )-1  \\
= \frac{1 + k/r + z_z(s_l-1)^2 + z_z (s_l-s)^2r/k}{\sqrt{(1 + k/r)^2 + z_z(1-s)^2}}-1.
\end{multline}
In the opposite limit $c/a \rightarrow 0$, we find instead
\begin{equation}
z_H^s \rightarrow z_z(s_l-s)^2\frac{r^2}{k^2} = \frac{\bar T(S_l - S_x)^2 \sigma_x}{\kappa_x}.
\end{equation}
In each of the limits, the sums cancel-out from the Harman FOM, leaving the results depending only on the anisotropy parameters.  Note that the argument used to derive these results did not depend on the contact locations on the $z=0$ face, or on the size of the current contacts.  All of the information about these factors is lost when the sums cancel-out.  In the thick sample limit, the FOM is a function of the parameters $z_z$, $s$, $s_l$ and $r/k$ only.  In the thin sample limit, the Harman FOM reduces to the FOM in the $x$-direction.  This is expected, since in the limit $c/a \rightarrow 0$, the FOM can only depend on the transport coefficients in the $x$-$y$-plane, which are isotropic.

In reality of course, one never has an infinitely thick or thin sample.  In deciding whether the system is in either regime, the magnitudes of $r$, $\mu_1$ and $\mu_2$ also need to be considered.  The sizes of the quantities $\pi \mu_i c/a$ and $\pi rc/a$ then determine whether or not the sample is in the thick/thin limit.  

\begin{table*}[h!]
\begin{ruledtabular}
\begin{tabular*}{1.0\textwidth}{>{\centering}m{1cm}<{} >{\centering}m{2cm}<{} c c  }
Case & Configuration &  Limit $c /a \rightarrow \infty$ & Limit $c /a \rightarrow 0$ \\ \hline
1  & \includegraphics[angle=180, width = 2cm]{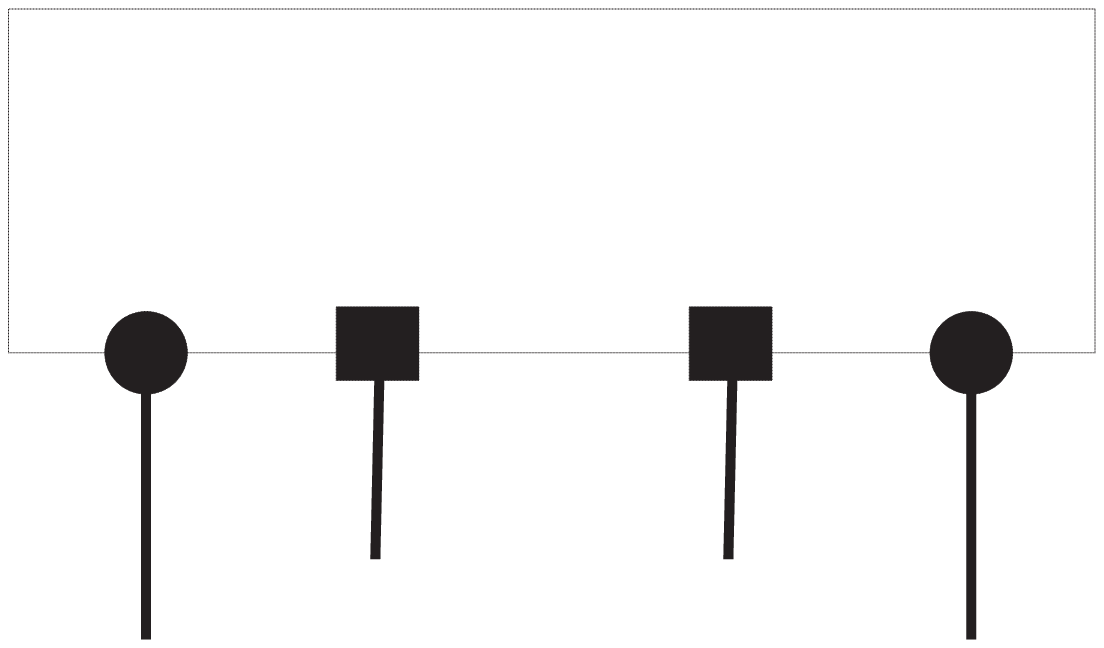}  & $z_H^s \rightarrow \displaystyle{\frac{1 + k/r + z_z(s_l-1)^2 + z_z (s_l-s)^2r/k}{\sqrt{(1 + k/r)^2 + z_z(1-s)^2}}-1}$ & $z_H^s  \rightarrow \displaystyle{z_z \frac {r^2}{k^2}(s_l-s)^2}$ \\ \hline
2 & \includegraphics[angle=180, width = 2cm]{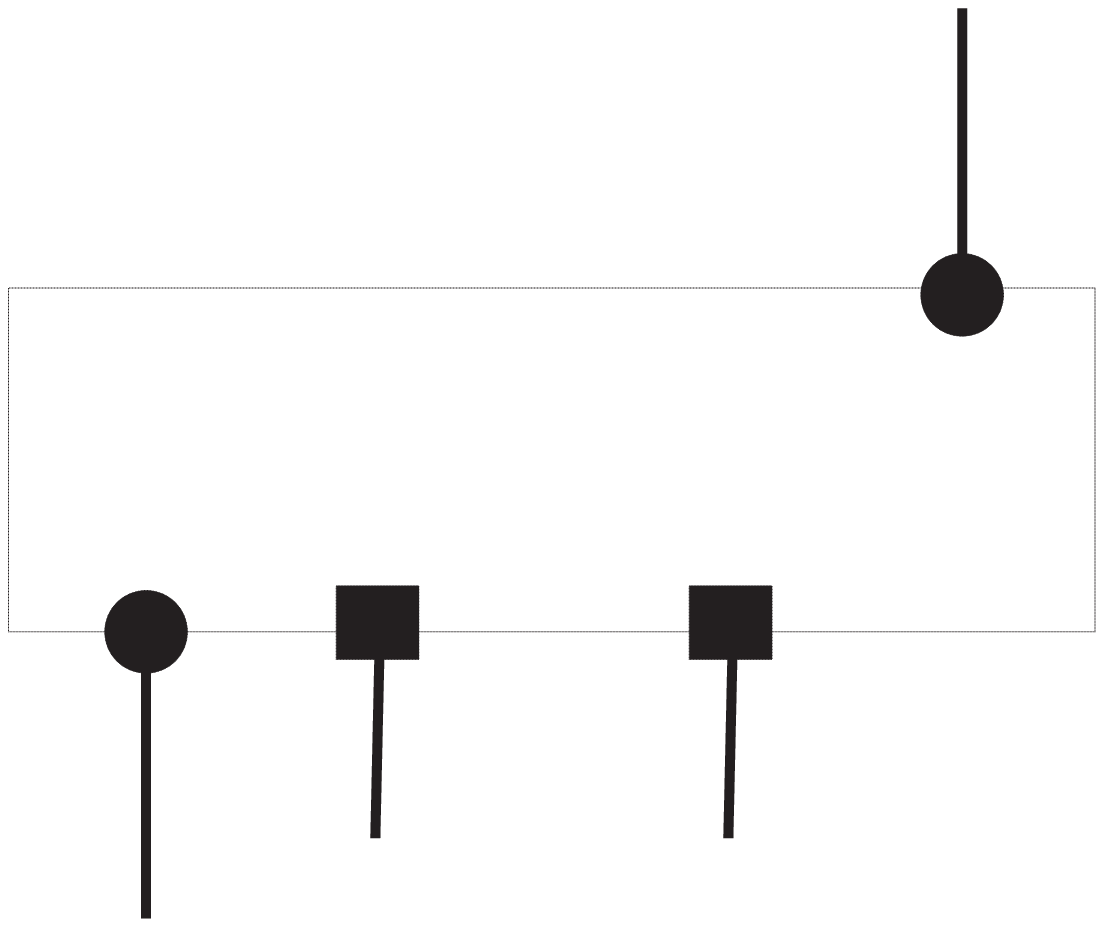} & $z_H^s \rightarrow \displaystyle{\frac{1 + k/r + z_z(s_l-1)^2 + z_z (s_l-s)^2r/k}{\sqrt{(1 + k/r)^2 + z_z(1-s)^2}}-1}$ & $z_H^s \rightarrow \displaystyle{z_z \frac {r^2}{k^2}(s_l-s)^2}$ \\ \hline
3 & \includegraphics[angle=180, width = 2cm]{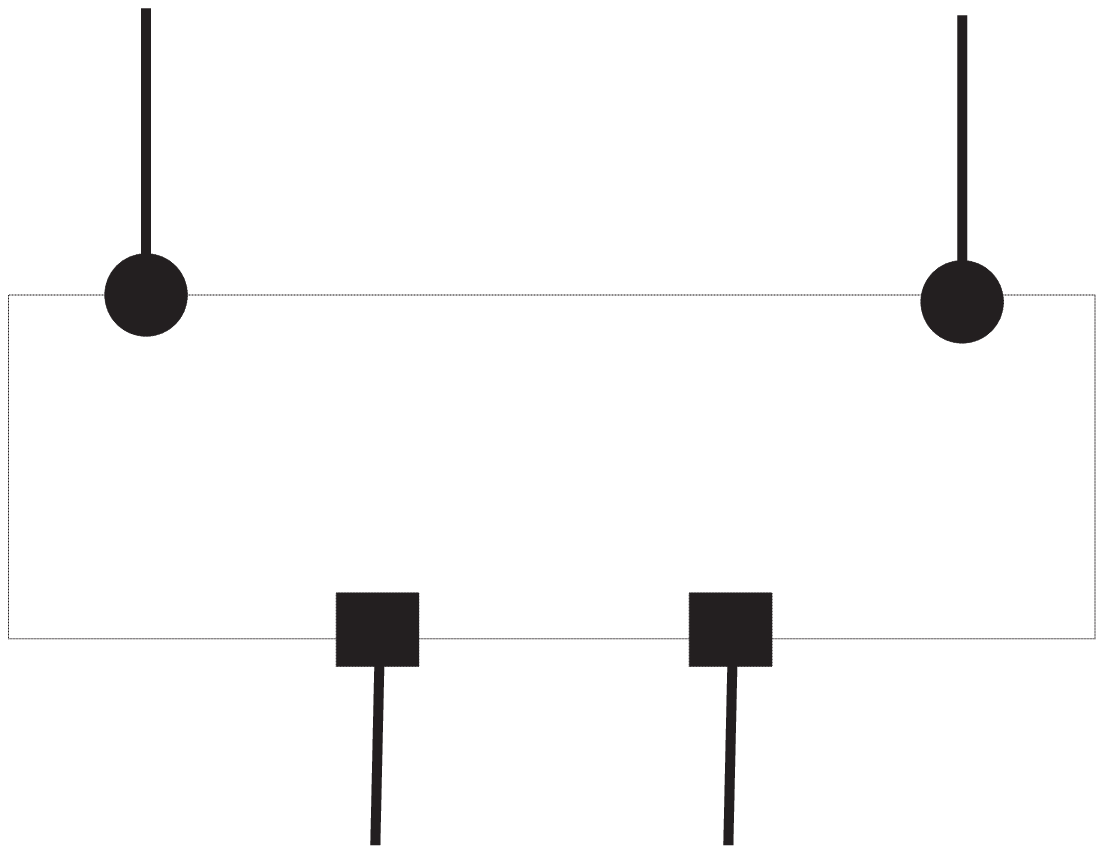} & $z_H^s \rightarrow \displaystyle{z_z \frac{r e^{\pi r c/a}}{\mu_2^2 - \mu_1^2} \sum_{i=1,2}(-1)^{i+1} \Big (\frac{\alpha'}{\mu_i} + \beta' \mu_i \Big )e^{-\pi \mu_i c/a}-1}$ & $z_H^s \rightarrow \displaystyle{z_z \frac {r^2}{k^2}(s_l-s)^2}$ \\ \hline
4 & \includegraphics[angle=180, width = 2cm]{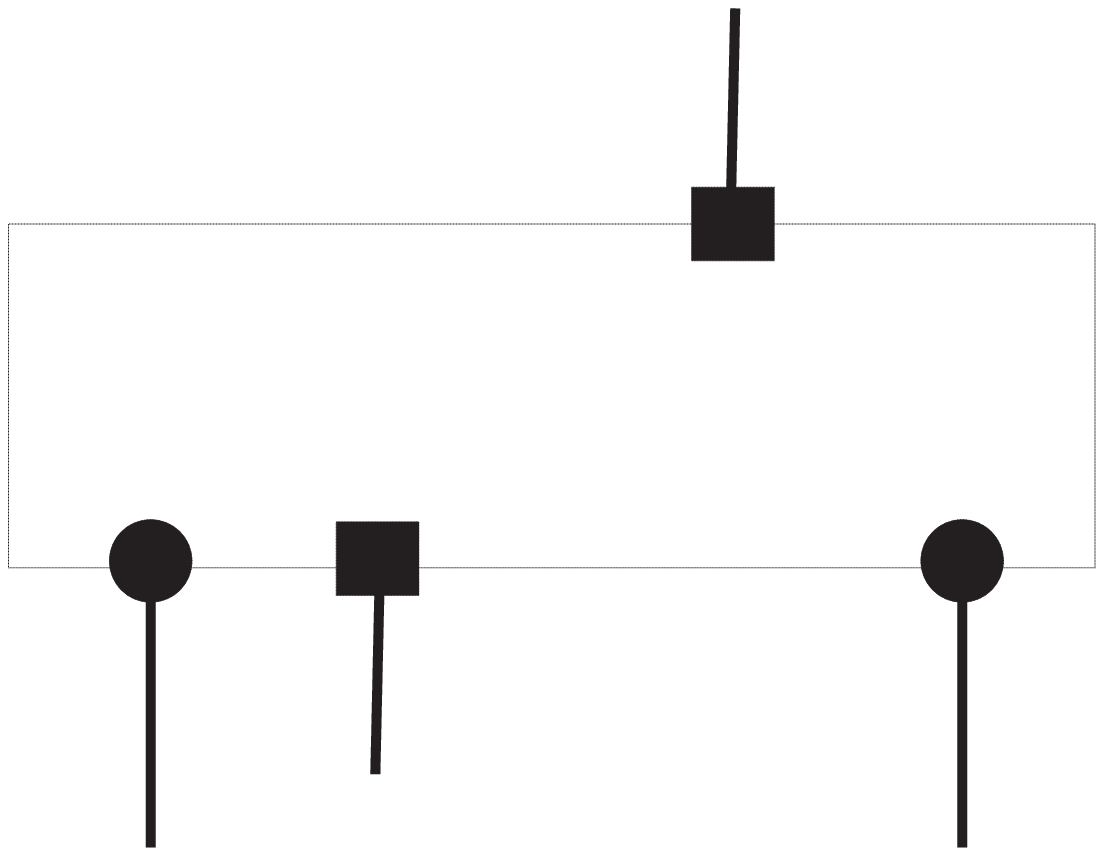}  & $z_H^o \rightarrow \displaystyle{\frac{1 + k/r + z_z(s_l-1)^2 + z_z (s_l-s)^2r/k}{\sqrt{(1 + k/r)^2 + z_z(1-s)^2}}-1}$ & $z_H^o  \rightarrow \displaystyle{z_z \frac {r^2}{k^2}(s_l-s)^2}$ \\ \hline
5 & \includegraphics[angle=180, width = 2cm]{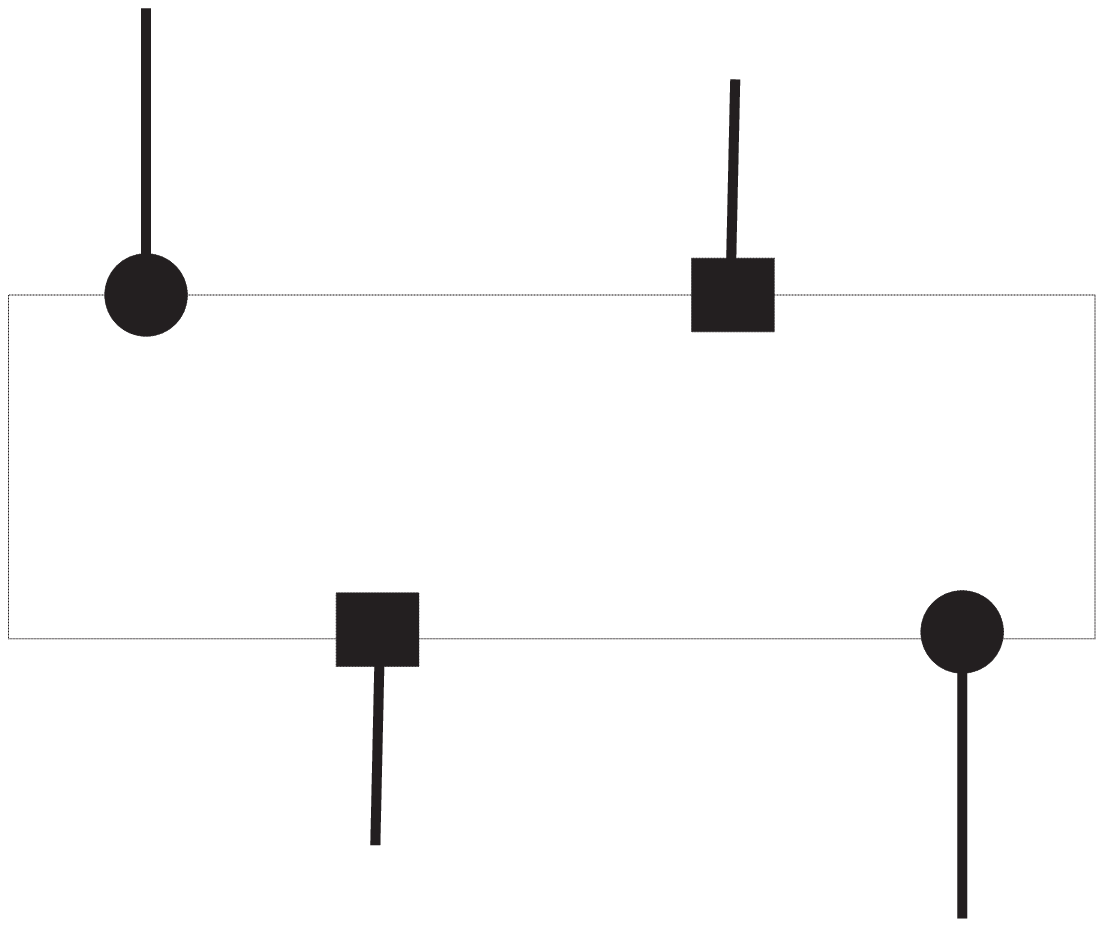} & $z_H^o \rightarrow z_z(s_l-1)^2$ & $z_H^o \rightarrow \displaystyle{z_z \frac {r^2}{k^2}(s_l-s)^2}$ 
\end{tabular*}
\end{ruledtabular}
\caption{\label{tab:same} Harman FOM for five different voltage and current contact configurations.  A schematic representation of the contact configuration is shown in the second column, where current contacts are indicated as squares, and the voltage contacts as circles.  The sample is orientated such that $z=0$ at the bottom.  Note that the actual locations of the contacts on the respective faces are arbitrary; the configurations shown are purely for illustration.}
\end{table*}

There are two other voltage contact locations which can be analysed within the general framework presented above.  We have compiled the results for the Harman FOM for these and the previous configuration in table \ref{tab:same}.   Case 2 turns out to be identical to case 1 in the thick/thin sample limits, but will give different results in an intermediate regime.  Case 3 is extremely surprising, since the Harman FOM is either exponentially large or small in the limit $c/a \rightarrow \infty$.  Again, all of these limits are independent of the voltage contact locations on the given faces, and of the size of the current contacts.  

A similar analysis can be applied to equation (\ref{eq:q2doppfom}) for current contacts on opposite faces of the sample.  Here, there are only two distinct voltage contact locations.  The results are also presented in table \ref{tab:same}.  Case 4 gives the same results as 1 and 3 in these limits.  More interesting is case 5, where the voltage contacts are on opposite sides of the sample.  In the thick sample limit, the summation terms tend to zero, leaving $z_H^o \rightarrow z_z(s_l-1)^2 = \bar T(S_l - S_z)^2 \sigma_z/\kappa_z$.  Hence in the limit that $c/a \rightarrow \infty$, the Harman FOM in this configuration coincides with the FOM in the $z$-direction, which would be measured in the conventional Harman method.  In the opposite limit of a thin sample, we instead find that the Harman FOM coincides with the FOM in the $x$-direction.

In fact, case 5 is just the small contact version of the conventional Harman method described in section \ref{sec:Harman}.  Our analysis suggests that the complications introduced by having a small current contact (or an imperfect, large contact) can be eliminated by making the sample very long and thin.  The size of the aspect ratio needed to do this depends on a number of factors.  As we pointed out previously, the particular regime the sample is in will also depend on the anisotropy through $r$, $\mu_1$ and $\mu_2$.  However, for the thick sample limit of case 5 other factors such as the contact locations and current contact size also play a major role.  The reason is the presence of the constant terms in equation (\ref{eq:q2doppfom}). For example, let us suppose that the sample thickness and anisotropy are such that we can make the previous $c/a \rightarrow \infty$ approximation to $\Delta \Omega_{mn}$.  Now, if the current contact is sufficiently small, and the voltage measured close enough to the current contacts, then overall the summation terms may still be larger than the constant term.  In this case, we would find that the Harman FOM would instead be given by equation (\ref{eq:samesamefom}).  For measurements further away from the current contacts, the summation terms may become small enough to recover the $c/a \rightarrow \infty$ result, $z_H \rightarrow z_z(s_l-1)^2$.  

The analysis above can be summarised as follows.  The thin sample limit of all five cases will result in a measurement of the in-plane FOM, $\bar T (S_l-S_x)^2\sigma_x/\kappa_x$.  The criteria for being in this limit are $\pi rc/a \ll 1$ and $\mu_irc/a \ll 1$, and in all cases the results only weakly depend on the contact locations and sizes.  In the thick sample limit (defined by $\pi rc/a \gg 1$ and $\mu_irc/a \gg 1$), cases 1, 2 and 4  depend only weakly on the contact locations and sizes, but the Harman FOM is a reasonably complicated function of the anisotropies and the intrinsic FOM.  Case 3 is also only weakly dependent on the contact locations, but gives an either exponentially large or small FOM, and so should be avoided.  

In principle case 5 is the best configuration, since it offers the possibility to measure the usual Harman FOM for the $z$-direction.  Roughly speaking, the criteria for being in the thick limit in this configuration with small contacts ($d^2/a^2 \ll 1$) are $a^3/cd^2\mu_i \ll 1$ and $a^3/cd^2r \ll 1$.  For large current contacts ($d^2/a^2 \lesssim 1$), we require instead that $(a-d)/\mu_ic \ll 1$ and $(a-d)/rc \ll 1$.  A compromise on the size of the contact would be needed, since it is also desirable to place the voltage contacts far away from the current contacts.  The reasons for this will become clear in section \ref{sec:case}.  An additional note of caution is that although a very long, thin sample will help reduce the errors associated with a small/imperfect contact, it may increase other sources of error in the experiment.  For example, errors due to heat loss via radiation would be increased by having a very long thin sample. 
 
\section{Limit of a Very Thin Sample; van der Pauw Formulae}

\begin{figure}[h!]
\begin{center}
\includegraphics[width=0.3\textwidth]{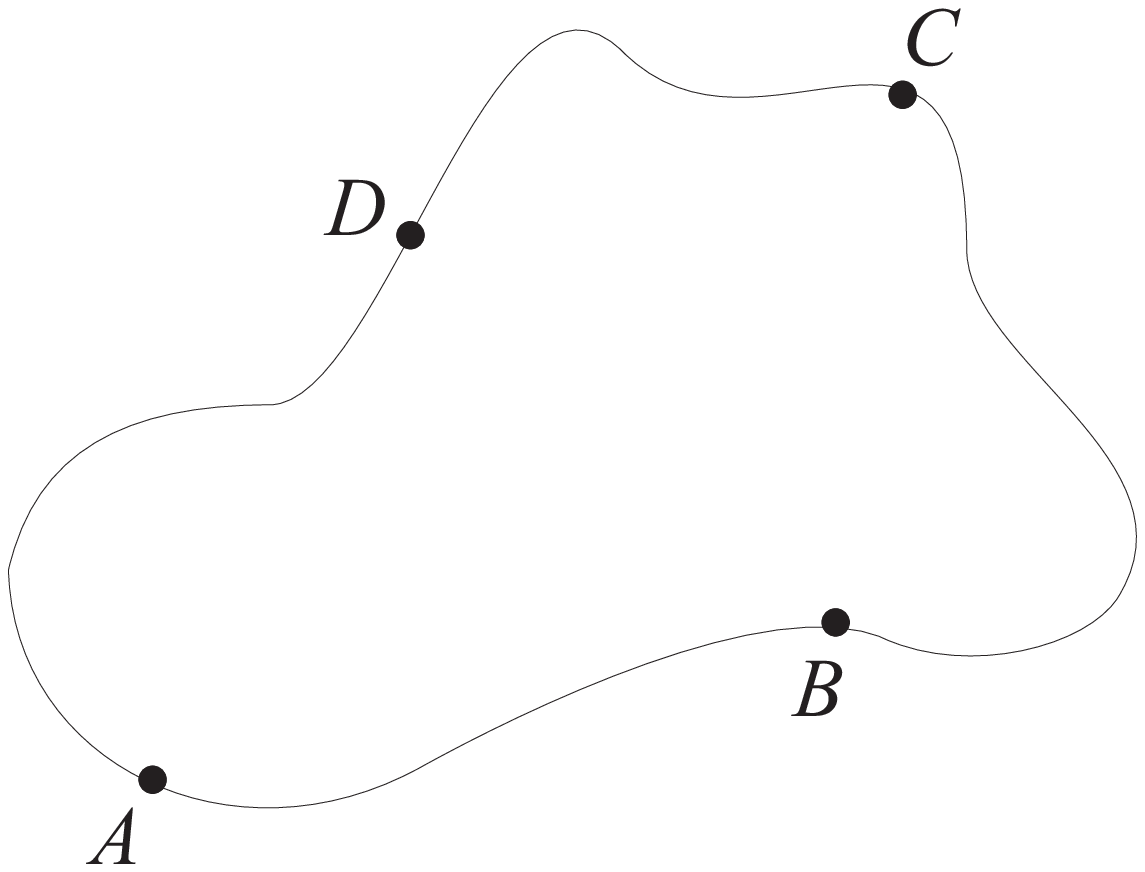}
\caption{\label{fig:Sample1}Schematic representation of the thin lamella to which van der Pauw's formula applies.  The point contacts labeled $A$, $B$ $C$ and $D$ are placed at arbitrary locations on the edge of the sample.} 
\end{center}
\end{figure}

In this section we take a brief diversion to consider an interesting application of the theory, which is unrelated to the Harman FOM.  Namely, we will prove the existence of generalised van der Pauw formulae for thin samples.  As discussed in the introduction, it has been shown by van der Pauw \cite{vdp1} that two four-terminal resistance measurements made with point contacts on a thin sample of arbitrary shape and isotropic resistivity obey a universal relationship.  Specifically, consider the sample shown in figure \ref{fig:Sample1}.  Then in the notation of section \ref{sec:gensoc}, the four-terminal resistance measurements $R_{ABCD} = V_{ABCD}/I_{AB}$ and $R_{BCDA} = V_{BCDA}/I_{BC}$ obey the equation
\begin{equation}\label{eq:vanderPauw}
\exp \Big (-\frac{\pi t}{\rho}R_{ABCD} \Big ) + \exp \Big (-\frac{\pi t}{\rho}R_{BCDA} \Big ) = 1,
\end{equation}
where $t$ is the thickness of the sample, and $\rho$ is the resistivity.

The proof of this result rests on the following.  In an isotropic system with point sources/sinks of electrical current (and neglecting thermoelectric effects), the electrical potential satisfies Poisson's equation (\textit{cf} equation (\ref{eq:eleccont2})).  It is well-known that the solutions of the two-dimensional Laplace equation (which is satisfied by the potential everywhere except at the points where the current is injected/removed) are invariant under conformal transformation.  Therefore, if the electrical potential satisfies a certain relationship in one sample geometry (\textit{i.e.} the van der Pauw formula), then this relationship continues to hold in any other sample geometry which can be produced by a conformal transformation.  In the original proof, van der Pauw used an infinite half-plane with contacts along the edge to establish equation (\ref{eq:vanderPauw}).   

Here, we will investigate whether analogous results hold for the temperature and voltage differences measured in a thin sample in the Harman set-up.  We will choose to let the sample thickness tend to zero along the $y$-direction, \textit{i.e.} $b \rightarrow 0$.  The van der Pauw formulae which we will derive involve the measured temperature and potential differences across the sample.  In order to properly include the thermopower of the leads, it is convenient to work with the quantity $\psi = \chi - S_l T = \chi - s_l\eta$, which we refer to as the `measured' voltage.   

In the thin sample limit, the measured voltage and temperature distributions satisfy the following equations  
\begin{widetext}
\begin{equation}\label{eq:vdp1}       
  r^2 \partwo{\psi}{x} + \partwo{\psi}z  + r^2 (s_l-s)\partwo{\eta}{x} +(s_l-1)\partwo{\eta}{z} = \frac{I}{\sigma_zb}\big (\delta(x-x_1)\delta(z-z_1) - \delta(x-x_2)\delta(z-z_2) \big ),
\end{equation}
\begin{equation}\label{eq:vdp2} 
 -\big (k^2z_z^{-1} - r^2s(s_l- s) \big)\partwo{\eta}{x} - (1+z_z^{-1}-s_l)\partwo{\eta}{z} + r^2s\partwo{\psi}{x} + \partwo{\psi}{z}   = \frac{s_lI}{\sigma_z b}\big (\delta(x-x_1)\delta(z-z_1) - \delta(x-x_2)\delta(z-z_2) \big ),
\end{equation}
\end{widetext}
which are easily derived from equations (\ref{eq:eleccont2}) and (\ref{eq:heatcont2}).  Because the measured voltage and temperature do not satisfy Poisson equations, the potential and temperature differences measured in a sample will not generally satisfy van der Pauw type formulae.  However, it turns out that it is possible to find linear combinations of $\psi$ and $\eta$ which \textit{do} satisfy uncoupled Poisson equations.  A convenient way of finding these linear combinations uses the solution for a rectangular geometry which we have found previously. Repeating the steps used in the derivation of (\ref{eq:intsameside}), we find the solution of (\ref{eq:vdp1}) and (\ref{eq:vdp2}) for a rectangular sample
\begin{multline}
\left(
\begin{array}{c}
\psi \\
\eta
\end{array}
\right) =  
\frac{I z_z}{\sigma_z abc} \sum_{i=1,2}'\frac{(-1)^{i+1}}{\mu_2^2-\mu_1^2}
\left(
\begin{array}{c}
\alpha' + \beta' \mu_i^2 \\
\epsilon + \delta \mu_i^2
\end{array}
\right) \\
\times \sum_{mn}\frac{\coss{\tm}{x}\coss{\tp}{z}}{\tp^2 + \mu_i^2\tm^2} \big (\coss{\tm}{x_2}\coss{\tp}{z_2} \\
- \coss{\tm}{x_1}\coss{\tp}{z_1} \big ) 
\end{multline}
Defining the following function,
\begin{multline}
\phi_i(x,z) = \frac{I z_z}{\sigma_z abc}
\sum_{mn}'\frac{\coss{\tm}{x}\coss{\tp}{z}}{\tp^2 + \mu_i^2\tm^2} \\
\times \big ( \coss{\tm}{x_2}\coss{\tp}{z_2} -\coss{\tm}{x_1}\coss{\tp}{z_1} \big ),
\end{multline}
$\psi$ and $\eta$ may be written as
\begin{equation}
\left (
\begin{array}{c}
\psi \\
\eta
\end{array}
\right )
=
\frac{1}{\mu_2^2-\mu_1^2}
\left (
\begin{array}{cc}
(\alpha' + \beta' \mu_1^2) &  -(\alpha' + \beta' \mu_2^2) \\
(\epsilon + \delta \mu_1^2) &  -(\epsilon + \delta \mu_2^2)
\end{array}
\right )
\left (
\begin{array}{c}
\phi_1 \\
\phi_2
\end{array} 
\right ). \nonumber
\end{equation}
This may be inverted to give $\phi_1$ and $\phi_2$ in terms of $\psi$ and $\eta$:
\begin{equation}\label{eq:transform}
\left (
\begin{array}{c}
\phi_1 \\
\phi_2
\end{array}
\right )
=
\frac{1}{\beta' \epsilon - \alpha' \delta}
\left (
\begin{array}{cc}
 -(\epsilon + \delta \mu_2^2) & (\alpha' + \beta' \mu_2^2) \\
-(\epsilon + \delta \mu_1^2) & (\alpha' + \beta' \mu_1^2)
\end{array}
\right )
\left (
\begin{array}{c}
\psi \\
\eta
\end{array}
\right ).
\end{equation}
Now, it is readily verified that $\phi_1$ and $\phi_2$ satisfy the following equation
\begin{multline}\label{eq:poissonlike}
\mu_i^2 \partwo{\phi_i}{x} + \partwo{\phi_i}{z} = \frac{I z_z}{\sigma_z b}\big (\delta(x-x_1)\delta(z-z_1) \\
- \delta(x-x_2)\delta(z-z_2) \big ).
\end{multline}
This can be written as a Poisson equation by rescaling the $x$-coordinate via $x = \mu_i \tilde x$ \textit{etc}:
\begin{multline}\label{eq:poisson}
\partwo{\tilde \phi_i}{\tilde x} + \partwo{\tilde \phi_i}{z} = \frac{I z_z}{\sigma_z b \mu_i}\big (\delta(\tilde x- \tilde x_1)\delta(z-z_1) \\
- \delta(\tilde x- \tilde x_2)\delta(z-z_2) \big ).
\end{multline}
Hence, the functions $\tilde \phi_1(\tilde x, z)$ and $\tilde \phi_1(\tilde x, z)$ are linear combinations of the measured voltage and temperature which satisfy uncoupled Poisson equations.  Note importantly that the parameters appearing in the transformation (\ref{eq:transform}) between $\phi_1$, $\phi_2$ and $\psi$, $\eta$ depend only on the transport coefficients, and not on the geometrical properties of the sample.  Indeed, if this relation is used to replace $\phi_i$ in equation (\ref{eq:poissonlike}), then it is straightforward to show that the result is simply a particular linear combination of equations (\ref{eq:vdp1}) and (\ref{eq:vdp2}).  Hence, we conclude that the transformation of equations (\ref{eq:vdp1}) and (\ref{eq:vdp2}) into (\ref{eq:poissonlike}) is independent of the rectangular geometry used in the derivation, and hence will hold for a sample of arbitrary shape.  In addition, since the normal derivatives of $\chi$ and $\eta$ vanish individually at the boundaries of the sample, so to do the normal derivatives of $\phi_i$.  Hence the boundary conditions on $\phi_i$ are the same as those for the temperature and voltage. 

The analysis of van der Pauw may now be applied directly to the fields $\tilde \phi_1$ and $\tilde \phi_2$.  This leads to van der Pauw formulae for differences in $\phi_i$ measured between points on the edge of a thin sample of arbitrary shape:
\begin{multline}
\exp \Big( \frac{\pi \sigma_z b\mu_i}{z_z I}\Delta \phi_{iABCD} \Big ) \\ 
+\exp \Big( \frac{\pi \sigma_z b\mu_i}{z_z I}\Delta \phi_{iBCDA} \Big )  =1.
\end{multline}     
In the general case, we have been unable to simplify these results any further.  In the limit that the thermopower becomes isotropic, $S_x=S_z=S$, these formulae reduce to 

\begin{multline}\label{eq:isovanderPauw1}
\exp \Big(\frac{\pi b\sqrt{\sigma_x \sigma_z}}{I}\big ( V_{ABCD} + (S_l-S)\Delta T_{ABCD} \big ) \Big) \\
+ \exp \Big(\frac{\pi b\sqrt{\sigma_x \sigma_z}}{I}\big ( V_{BCDA} + (S_l-S)\Delta T_{BCDA} \big ) \Big) = 1,
\end{multline}
\begin{multline}\label{eq:isovanderPauw2}
\exp \Big(-\frac{\pi b\sqrt{\kappa_x \kappa_z}}{(\Pi_l-\Pi) I}\Delta T_{ABCD} \Big) \\
+ \exp \Big(-\frac{\pi b\sqrt{\kappa_x \kappa_z}}{(\Pi_l-\Pi) I}\Delta T_{BCDA} \Big) = 1.
\end{multline}
The combination $V + (S_l - S)\Delta T$ is recognised as the ac voltage, $V_{\mathrm {ac}}$.  Hence, the first of these equations reduces to the classic form (\ref{eq:vanderPauw}) of van der Pauw's formula, where the resistivity is replaced by the geometric mean resistivity, $(\sigma_x\sigma_z)^{-1/2}$.  Note that the apparent sign difference between (\ref{eq:isovanderPauw1}) and (\ref{eq:vanderPauw}) is due to the fact that $V$ and $V_{\mathrm{ac}}$ are electrochemical potential differences, rather than electrical potential differences.  Equation (\ref{eq:isovanderPauw2}) clearly represents the direct thermal analogue of the classic van der Pauw formula, where the combination $\Delta T/(\Pi_l - \Pi)I$ is just the thermal resistance between the measurement points.   

Finally, we note that the results of other works \cite{Perloff, Bergemann} on the voltage and current distributions in thin samples may also be applied to the thermoelectric problem.  One simply replaces the electrical potential with the field $\tilde \phi_i$, and then uses the transformation (\ref{eq:transform}) to relate the results to the actual measured voltage and temperature in the sample.

\section{Application to Experiment}\label{sec:case}

\subsection{Results}

Having established all of the basic results, we now consider an application of the theory to some experiments.  Recently, Kobayashi \textit{et al} have developed a `modified' Harman method, and used it to measure the figures of merit of several anisotropic materials \cite{Kobayashi, Tamura}.  The modification to the usual Harman method divides into two parts.  The first concerns the way in which the ac and dc voltages are measured, and does not concern us here.  The second modification is that small current contacts are used, and it is their effect that we wish to analyse using the formalism developed in this paper.

\begin{figure}[h]
\begin{center}
\includegraphics[width=0.35\textwidth]{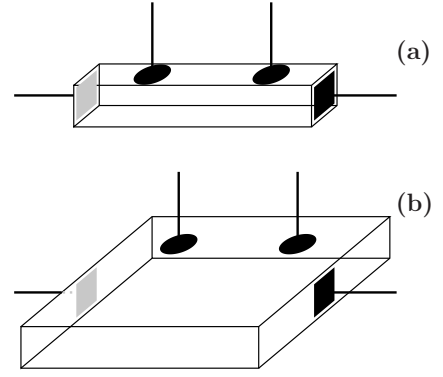}
\caption{\label{fig:Sample3a}Schematic representations of the two sample configurations used in Kobayashi \textit{et al's} experiments.  The current contacts are shown as squares, while the voltage contacts are indicated by circles.  (a) Configuration for  $\mathrm{Ba}_{1.2}\mathrm{Rh}_8\mathrm{O}_{16}$, with the current directed along the $b$-axis  (b) Configuration for $\mathrm{Bi}_{0.78}\mathrm{Sr}_{0.4}\mathrm{O}_{3+d}$ and \bico, with the current directed along the $a$-$b$ plane.} 
\end{center}
\end{figure}

We begin with three cases where the modified Harman method seems to work well, \textit{i.e} the Harman FOM obtained is physically reasonable, and in one case is known to be consistent with values from other measurements.  In the first experiment, the material is $\mathrm{Ba}_{1.2}\mathrm{Rh}_8\mathrm{O}_{16}$, a quasi-one dimensional rhodium oxide.  Figure \ref{fig:Sample3a}(a) shows the sample and contact configuration used in the measurement.  The sample is orientated such that the current flows along the $b$-axis of the material.  The key features of this configuration are the small contact face ($a=b=0.05$mm) and large aspect ratio ($c/a=10$).  Since the current contacts extend almost completely over the contact faces, the flows of heat and electricity are essentially uniform.  Hence the Harman FOM is expected to coincide with the intrinsic FOM in along the $b$-axis.

The sample configurations in the other two cases are very similar, and are shown schematically in figure \ref{fig:Sample3a}(b).  Crystals of two different layered oxides, $\mathrm{Bi}_{0.78}\mathrm{Sr}_{0.4}\mathrm{O}_{3+d}$ and BiPbSrCoO, are used, with the current applied parallel to the $a$-$b$ plane.  Both crystals have very small aspect ratios, $c/a=$0.02-0.03. The layered structure means that the largest anisotropy occurs between in and out-of-plane directions, with the in-plane directions relatively isotropic.  Although we have not considered this particular combination of sample configuration and anisotropy previously, it is straightforward to understand why the Harman FOM coincides with the intrinsic in-plane FOM.  First, because the sample is very thin, the current contact extends almost completely over the full thickness of the sample.  This means that the temperature and voltage distributions do not significantly vary along the out-of-plane direction.  They do however vary along the in-plane directions, due to the current contact being much smaller than the width of the sample.  This leads to non-uniform flows of heat and electricity in the in-plane directions.  However, the transport properties are only weakly anisotropic between the in-plane directions, and hence heat and electricity flow along very similar paths.  This leads to a situation similar to that described at the end of section \ref{sec:hfom} B, where the effects of the small current contact cancel-out between the measured electrical resistance and thermal conductance.  Hence, the Harman method also works well in this particular configuration.       

In addition to these examples where the Harman method works well, two other configurations have been investigated by Tamura \textit{et al}, where the Harman FOM is found to substantially larger then the relevant intrinsic FOM.  Both experiments were performed on BiPbSrCoO single crystals, with the contacts faces parallel to the $a$-$b$ plane.  To understand the reason for increased Harman FOM, we first need a more detailed description of the anisotropy in BiPbSrCoO. 

One of the most extraordinary features of this compound is the degree of anisotropy in the electrical resistivity between in-plane and out-of-plane directions.  The out-of-plane resistivity is around $10^4$ times larger than the in-plane at room temperature, with the degree of anisotropy increasing still further at lower temperatures.  The in-plane resistivity of BiPbSrCoO is a few m$\Omega$cm at room temperature \cite{Fujii}, making it a relatively poor conductor of electricity even in the in-plane direction.  The large anisotropy with respect to charge transport is not seen in the thermal conductivity, with the crystal structure making the in-plane thermal conductivity larger than the out-of-plane by around 7 at room temperature \cite{Terasaki3}.  A final remarkable property of this compound is that despite the large electrical anisotropy, the in-plane thermopower is only larger than the out-of-plane by a factor of 2 at room temperature.  At room temperature, the dimensionless in-plane FOM $z_{ab} = 0.029$, while the large electrical resistivity gives a very small out-of-plane FOM, $z_c = 9.72 \times 10^{-6}$ at the same temperature.  

\begin{figure}[h]
\begin{center}
\includegraphics[width=0.45\textwidth]{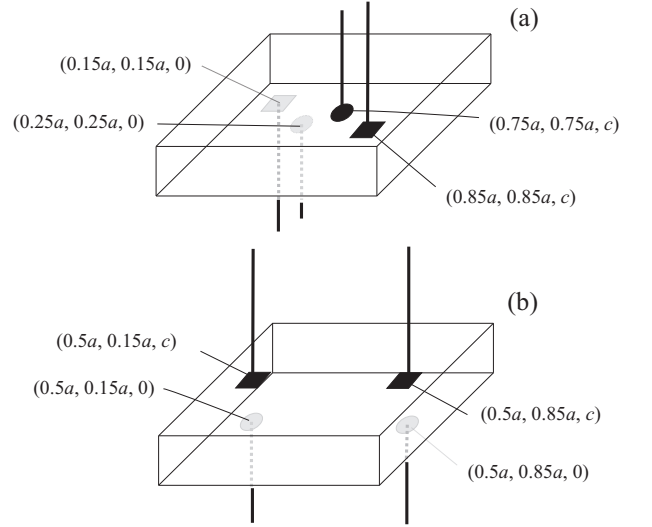}
\caption{\label{fig:Sample4}Schematic representations of the two sample configurations used in Tamura \textit{et al's} experiments.  The coordinate system is the same as that used in figure \ref{fig:3dsample}.  (a) Current contacts on opposite sides of the sample.  (b) Current contacts on the same side of the sample.} 
\end{center}
\end{figure}

Since the material anisotropy is of quasi-two-dimensional form, we make use of the results of section \ref{sec:q2d1}.  The sample is orientated such that the $c$-axis is aligned in the $z$-direction, with the $a$ and $b$-axes lying along $x$ and $y$ respectively.  The sample has dimensions in the $x$, $y$ and $z$-directions of $W\times H \times L$, with $W=2$mm, $H=2.1$mm and $L=0.45$mm.  All lengths will be measured relative to $W$, giving a sample with $a\approx b=1$, and $c=0.25$. The first contact configuration considered is shown schematically in figure \ref{fig:Sample4}(a).  The current contacts with a width of approximately $0.065a$ are located at $(0.15a, 0.15a, 0)$ and $(0.85a, 0.85a,c)$, while the voltage contacts are located at $(0.25a, 0.25a, 0)$ and $(0.75a, 0.75a, c)$.  At $180$K, the anisotropy parameters of the material are $r \approx 140$, $k \approx 2.8$ and $s \approx 1.9$.  The out-of-plane dimensionless FOM, $z_z$, is $1.35 \times 10^{-5}$ at the same temperature.  In the following we will neglect the thermopower of the current and voltage leads, which are much smaller than that of the sample in either direction.  From our theory, the ac resistance is this configuration is predicted to be $R_{\mathrm{ac}} = 1.05R_{z0}$, while the thermoelectric resistance is $R_{\mathrm{dc}} - R_{\mathrm{ac}} = 12.7\bar TS_z^2/K_{z0}$.  This predicts the measured Harman FOM to be approximately 12$z_z$.  

\begin{figure}[h]
\begin{center}
\includegraphics[width=0.45\textwidth]{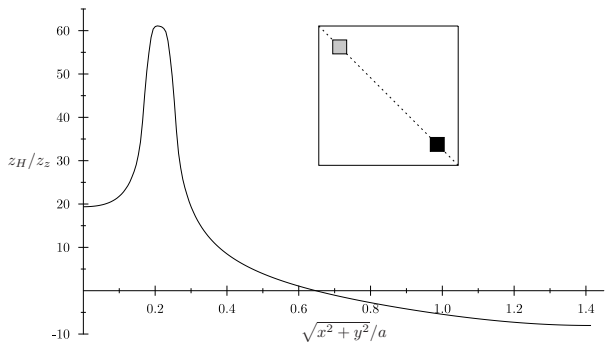}
\caption{\label{fig:tempcut}Harman FOM measured between points $(x,y,0)$ and $(a-x,b-y,c)$, for $x$ and $y$ lying on the dotted line indicated in the inset.  In the experiment the centers of the voltage contacts lie at $(0.25a,0.25a,0)$ and $(0.75a,0.75a,c)$.} 
\end{center}
\end{figure}

Experimentally, the ac resistance, thermoelectric resistance and Harman FOM at 180K are measured to be $1.6R_{z0}$, $57 \bar TS_z^2/K_{z0}$ and $35.5z_z$ respectively.  The reason for poor agreement between theory and experiment is not completely clear, but may be partly due to the contact configuration employed in this measurement.  First, the choice of the constant flux boundary condition is expected to produce the largest errors close to the current contacts, which is where the voltage measurements are made.  Second, as we pointed out in the last section, this particular contact configuration makes the Harman FOM extremely sensitive to the exact positions and sizes of the current contacts.  To highlight this, we plot in figure \ref{fig:tempcut} the FOM measured between points $(a-x,a-y,c)$ and $(x,y,c)$, for $x$ and $y$ lying on the line indicated in the figure.  The maximum value of the FOM occurs between the centres of the current contacts, and is around 60$z_z$.  It is clear that the FOM varies extremely rapidly in the vicinity of the voltage contact.  The finite size of the voltage contact will also mean that the measured FOM will be some average of the region beneath the contact, which has not been accounted for here.

The second configuration used in the experiment is shown in figure \ref{fig:Sample4}(b).  As we pointed out in section \ref{sec:thickthin}, in this configuration the Harman FOM is only weakly dependent on the contact sizes and locations on the contact faces.  In fact, the sample turns out to be approximately in the thick sample limit, since $\pi \mu_1 c/a \approx \pi r c/a \approx 110$ and $\pi \mu_2 c/a \approx \pi k c/a \approx 2.2$.  We can therefore use the result for case 3 in table \ref{tab:same}, which reduces to   
\begin{equation}
z_H^s \approx z_z s^2\frac{r}{k}e^{\pi(r-k)c/a}.    
\end{equation}
The exponential factor means that the Harman FOM is expected to be enormous, with $z_H \approx 1.2 \times 10^{49} z_z$!  This result occurs because the large electrical anisotropy causes the electrical current to flow in a very thin layer (with width of order $a/r$) near to the contact face at $z=0$ \cite{Busch}.  The ac potential difference measured between points on the bottom of the sample at $z=c$ is therefore exponentially suppressed, with the theory predicting that $R_{\mathrm{ac}} = 2 \times 10^{-49}R_{z0}$.  In contrast, the heat current penetrates the sample over a distance of the order of $a/k$, and so the temperature difference between the voltage contacts is relatively large, with the thermoelectric resistance predicted to be $R_{\mathrm{dc}} - R_{\mathrm{ac}} = 2.48 \bar T S_z^2/K_{z0}$.    

Experimentally the Harman FOM in this configuration is extremely large, although not to the degree predicted by the theory.  At 180K, the FOM is measured to be around 5.5, or equivalently $4 \times 10^{5}z_z$, which is still an extremely large enhancement of the intrinsic FOM.  The measured ac resistance is $1.1 \times 10^{-4}R_{z0}$, although the origin of this resistance is unclear given that the theory predicts that the ac voltage should be effectively zero in this configuration. Also puzzling is the magnitude of the measured thermoelectric resistance,  $R_{\mathrm{dc}} - R_{\mathrm{ac}} = 44.2 \bar T S_z^2/K_{z0}$, which is comparable to that measured in the previous configuration.                         

In summary, we have used the results derived in section \ref{sec:q2d1} to qualitatively understand the enhanced Harman FOM measured by Tamura \textit{et al} in BiPbSrCoO.  In both configurations, we find that the theory underestimates the sizes of the ac and thermoelectric resistances in the sample.  One possible interpretation of the results is that the sample is behaving as if the effective electrical and thermal anisotropies were considerably \textit{smaller} than the values expected from the resistivities and thermal conductivities.  Alternatively, the uncertainties associated with the current and voltage contact locations could also explain some of these results.  

\subsection{Physical Origin Of the Enhanced Figure of Merit}       

In this section we will attempt to understand the physical origin of the enhanced FOM predicted by the theory in first sample configuration investigated by Tamura \textit{et al}.  There are two distinct processes at work.  The first concerns the ways in which heat and electricity flow through the sample, while the second involves the anisotropic thermoelectric effects.  Let us assume temporarily that the thermopower is isotropic.  In the first contact configuration the ac voltage and temperature distributions are given by equations (\ref{eq:acoppside}) and (\ref{eq:tempoppside1}) respectively, with $R_{mn}=r\omega_{mn}$ and $K_{mn}=k\omega_{mn}$.  Evaluating the ac potential difference between the voltage contacts gives an ac resistance of 1.05$R_{z0}$, while the measured thermal resistance between the same two points is $4.28K_{z0}^{-1}$.  Hence the measured thermal resistance is increased by a much larger factor than the electrical resistance.  The reason for this is the large difference in the electrical and thermal anisotropy, which causes electricity and heat to flow along very different paths through the sample (\textit{cf} figure \ref{fig:flows}).  

From equation (\ref{eq:isofom}) (and taking $S=S_z$), the Harman FOM is evaluated to be 3.8$z_z$.  Given that this enhancement is less than the factor of 12 from evaluating equation (\ref{eq:q2doppfom}), the remaining enhancement must come from the effects of thermoelectric anisotropy.  To understand these effects, we will make some approximations to the formulae derived in section \ref{sec:q2d1}.  The large electrical anisotropy and small intrinsic FOM suggest that we take the combined limit $r \gg 1$ and $z_z \ll 1$.  To leading order, the quantities $\mu_1$ and $\mu_2$ are given by
\begin{align}  
\mu_1^2 &\approx r^2 \big (1 + z_z(1-s)^2 \big ),  \nonumber\\
\mu_2^2 &\approx k^2 \big (1 - z_z(1-s)^2 \big ).  \nonumber
\end{align}
Substituting these expressions into equations (\ref{eq:q2doppside}), and making the $r \gg 1$ and $z_z \ll 1$ approximations, we find for current contacts on opposite sides of the sample:
\begin{multline}\label{eq:largervoltoppside}
\chi^o(x,y,z)= \frac{Ic}{\sigma_z ab}\Big ( \frac{z}{c} - \frac{1}{2} \Big)\Big ( 1 - \frac{\sigma_z S_z(\Pi_l - \Pi_z)}{\kappa_z}\Big) + \\
\frac{I}{\sigma_z d^2}\bigg( \Big(1 - \frac{\sigma_z(S_z-S_x)}{\kappa_z}(\Pi_l - \Pi_z - \Pi_x)\Big)\sum_{mn}'\Omega_{mn}(x,y,z;r) \\
-S_x\frac{(\Pi_l - \Pi_x)}{\kappa_z}\sum_{mn}'\Omega_{mn}(x,y,z;k)\bigg),
\end{multline}
\begin{multline}\label{eq:largertempoppside}
T^o(x,y,z)=  -\frac{I(\Pi_l - \Pi_z)c}{\kappa_z ab}\Big ( \frac{z}{c} - \frac{1}{2} \Big)  + \\
\frac{I}{\kappa_z d^2}\Big((\Pi_z - \Pi_x)\sum_{mn}'\Omega_{mn}(x,y,z;r) \\
-(\Pi_l - \Pi_x)\sum_{mn}'\Omega_{mn}(x,y,z;k)\Big).
\end{multline}
It is instructive to compare these expressions with those for isotropic thermopower, equations (\ref{eq:acoppside}) and (\ref{eq:tempoppside1}).  The temperature distribution now has an additional term, a direct result of the Bridgman effect which allows the current distribution \textit{within} the the sample to alter the temperature distribution.  This term is very small in this case since it is proportional to $1/r$.  The third term is also present in equation (\ref{eq:tempoppside1}), but in this limit it is the \textit{in-plane} Peltier coefficient which appears in the prefactor.  This is quite surprising, given that because the current contacts are placed on the faces at $z=0,c$, and the current enters the sample along the $z$-direction!  The large electrical anisotropy combined with the Bridgman effect are responsible for this.  Let us (trivially) rewrite the second the term in $T^o$ as 
\begin{multline}
(\Pi_l - \Pi_x)\sum_{mn}'\Omega_{mn}(x,y,z;k) = \\
\big ( (\Pi_l - \Pi_z) + (\Pi_z - \Pi_x) \big ) \sum_{mn}'\Omega_{mn}(x,y,z;k),
\end{multline}
which we interpret in the following way.  The current enters the sample in the $z$-direction, causing a Peltier effect proportional to $(\Pi_l - \Pi_z)$, which is captured by the first term.  Due to the large electrical resistivity in the $z$-direction, the electrical current turns to flow more easily in the $x$-direction, as shown in figure \ref{fig:flows}(a).  This generates a Bridgman effect proportional to $(\Pi_z - \Pi_x)$, which we associate with the second term.  In the limit that $r \rightarrow \infty$, this process occurs right at the current contact, and therefore it is as if the current flows directly from the leads into the $x$-direction.  The reverse process happens at the other contact.  Hence the relative Peltier coefficient is $(\Pi_l - \Pi_x)$.  Note that prefactors of the linear terms still involve the out-of-plane thermoelectric coefficients, since the uniform flows of heat and electricity resulting from them are unaffected by the Bridgman effect.

The voltage distribution is also modified by the thermoelectric anisotropy. The term proportional to $(S_z - S_x)$ is interpreted as the Ohmic contribution from the thermoelectric eddy currents in the sample.  Again, this term is proportional to $1/r$ and is small in this case.  The most important feature of the voltage distribution is that the final thermoelectric term is proportional to the in-plane thermopower.  

The net effect of this is that because the in-plane thermoelectric coefficients appear in the dominant term of equation (\ref{eq:largervoltoppside}), the FOM has an enhancement due to thermoelectric effects of roughly $s^2 = 3.6$.  This approximately accounts for the difference between the figures of merit evaluated from equations (\ref{eq:q2doppfom}) and (\ref{eq:isofom}).  The actual enhancement is less than this, because the the sample is not strictly in the $r \rightarrow \infty$ limit.

\section{Effective figure of merit}\label{sec:efffom}

The results of the previous sections have indicated that the effects of the small current contact and anisotropy can lead to the Harman FOM being larger than the intrinsic FOM of the material in the direction perpendicular to the contact face.  In the second contact configuration investigated by Tamura \textit{et al}, the measured increase in the FOM was extremely large, with the theory predicting an exponential increase in this case.  However, the enhancement only occurs for a certain set of voltage contact locations; there are others for which the Harman FOM is actually reduced (see figure \ref{fig:tempcut}).  Hence, it is not at all clear whether the increased Harman FOM indicates an increase in the thermoelectric performance of this system, or is simply an artefact of the measurement.  It is this question that we shall attempt to address here.

The central problem with the Harman FOM as defined is that it is a \textit{local} quantity, depending on voltages measured between particular points in the sample.  In order to decide whether the thermoelectric performance is improved, we need to define a FOM which characterises the properties of the \textit{whole} sample.  This can be achieved through a consideration of the rate of irreversible entropy production in the system.  The concept of entropy production as applied to thermoelectric devices is not a new idea.  Sherman \textit{et al} \cite{Sherman} were the first to point-out that maximising the efficiency of a thermoelectric device is `equivalent' to minimising the rate of unwanted entropy production.  Specifically, they considered a generator made from a thermoelectric material with temperature-independent transport coefficients connected to an external load. It was demonstrated that the load resistance which maximises the efficiency also minimises the rate of entropy production in the generator divided by the rate of entropy production in the load.  More recently, Snyder \cite{Snyder} has introduced the thermoelectric compatibility as the quantity which maximises the local efficiency of a generator.  This quantity may be derived by considering the local rates on entropy production.

Within the framework of irreversible thermodynamics \cite{deGroot}, the rate of entropy production is written in terms of pairs of conjugate forces and fluxes.  For the thermoelectric problem, we choose the fluxes to be the electrical and heat currents $\mathbf j$ and $\mathbf j_Q$, for which the appropriate (linearised) conjugate forces are $(1/\bar T)\nabla \chi$ and $-(1/\bar T^2)\nabla T$.  The rate of entropy production per unit volume is given by   
\begin{equation}
\dot{\mathcal S} = \frac{j_i}{\bar T} \parone{\chi}{x_i} - \frac{j_{Qi}}{\bar T^2}\parone{T}{x_i}.
\end{equation}
Using equations (\ref{eq:anisoheat}) and (\ref{eq:anisochi}) to eliminate $j_{Qi}$ and $\textparone{\chi}{x_i}$ from these expressions leads to 
\begin{align}
\dot{\mathcal S} = \frac{1}{\bar T}\rho_{ik}j_ij_k + \frac{\kappa_{ik}}{\bar T^2}\parone{T}{x_i}\parone{T}{x_k},
\end{align} 
where the interference term proportional to $j_k \textparone{T}{x_i}$ vanishes due to the Kelvin/Onsager relation $\Pi_{ik} = \bar T S_{ki}$.  The first term in this expression is the entropy production due to Joule heating, while the second is the entropy production associated with heat flow.  The total rate of entropy production is then the integral of this expression over the volume:
\begin{equation}  
\Sigma = \int dV \dot{\mathcal S} = \Sigma_J + \Sigma_T, 
\end{equation}
where $\Sigma_J$ and $\Sigma_T$ are the total rates of entropy production due to Joule heating and heat flow respectively. 

To gain some intuition about $\Sigma$, let us evaluate it in the conventional Harman method.  Using equations (\ref{eq:1dtemp}) and (\ref{eq:1dchi}), we find the total rate of entropy production is given by
\begin{align}\label{eq:totalent}
\Sigma &= \Sigma_J + \Sigma_T = \frac{I^2 R_{z0}}{\bar T} + \frac{(S_l-S_z)^2I^2}{K_{z0}} \nonumber \\
&= \frac{I}{\bar T}\big ( V_{\mathrm{ac}} - (S_l - S_z)\Delta T \big ) = \frac{IV_{\mathrm{dc}}}{\bar T}. 
\end{align} 
Hence the total rate of entropy production is simply related to the power obtained from the measured (two terminal) voltage and the current.  The key observation from this analysis is that the FOM of the sample may be written in terms of $\Sigma_T$ and $\Sigma_J$:
\begin{equation}\label{eq:zeff}
z = \frac{\Sigma_T}{\Sigma_J} = \frac{\bar T (S_l - S_z)^2}{R_{z0}K_{z0}}.
\end{equation}
The advantage of this expression compared to the Harman definition (\ref{eq:figofmerit}) is that $\Sigma_T$ and $\Sigma_J$ are quantities which take single values for a sample with a particular configuration of current contacts.  The freedom with respect to the locations of the voltage contacts has been eliminated.

Let us now apply this result in the regime of small current contacts and isotropic thermopower.  In appendix \ref{sec:ent} we show that equation (\ref{eq:totalent}) continues to hold, but where the potential and temperature differences are replaced by \textit{contact averaged} values.  The contact averaged potential/temperature difference is defined as the average potential/temperature difference between points lying on the current contacts.  The rate of entropy production due to Joule heating is related to the contact averaged ac voltage via $\bar T \Sigma_J = I\langle V_{\mathrm{ac}} \rangle$.  We can therefore use this relation to define an effective electrical resistance for the sample, $R_{\mathrm{eff}} = \langle V_{\mathrm{ac}} \rangle/I = \bar T \Sigma_J/I^2$.  This is the resistance which is `seen' by the external voltage source which drives the current through the sample.  Similarly, the rate of entropy production due to heat flow is related to the contact averaged temperature difference across the sample via $\bar T \Sigma_T = -(S_l-S)I\langle \Delta T\rangle$.  Hence an effective thermal conductance can be defined as $K_{\mathrm{eff}} = -(\Pi_l-\Pi)I/\langle \Delta T \rangle = \bar T^2 \Sigma_T/\langle \Delta T \rangle^2$.  Finally, the effective FOM is given by
\begin{align}    
&z_{\mathrm{eff}} = \frac{\Sigma_T}{\Sigma_J} = \frac{\bar T (S_l-S)^2}{R_{\mathrm{eff}}K_{\mathrm{eff}}}. \nonumber 
\end{align}
The effective FOM is related to an effective electrical resistance and thermal conductance, which reflect the flows of heat and electricity through the whole sample.

The true value of defining the FOM in terms of the entropy production in the Harman set-up is only seen when the thermopower becomes anisotropic.  For anisotropic thermopower, the individual rates of entropy production due to Joule heating and heat flow are not formally related to the contact averaged ac potential difference and temperature difference respectively.  However, the relation between the \textit{total} rate of entropy production and the contact averaged dc potential difference, $\bar T \Sigma = I \langle V_{\mathrm{dc}} \rangle$, \textit{does} still hold.  The problem previously was that we did not know how to divide the dc voltage into an Ohmic part and a thermoelectric part; the rate of entropy production shows how this division should be made.  If we \textit{define} $\langle V_{\mathrm{dc}} \rangle = IR_{\mathrm{eff}} + S_{\mathrm{eff}}\Delta T$, and $\Delta T = \bar T S_{\mathrm{eff}}/K_{\mathrm{eff}}$, then the effective resistance, thermopower, thermal conductance and FOM are \textit{defined} by
\begin{eqnarray}
R_{\mathrm{eff}} = \frac{\bar T \Sigma_J}{I^2}, & \displaystyle{S_{\mathrm{eff}} = \frac{\bar T \Sigma_T}{I\langle \Delta T \rangle}}, & \displaystyle{K_{\mathrm{eff}} = \frac{\bar T^2 \Sigma_T}{\langle \Delta T \rangle^2}} \nonumber \\
~ & \displaystyle{z_{\mathrm{eff}} = \frac{\Sigma_T}{\Sigma_J} = \frac{\bar T (S_{\mathrm{eff}})^2}{R_{\mathrm{eff}}K_{\mathrm{eff}}}}. & ~ \nonumber 
\end{eqnarray}
where $\Sigma_J$ and $\Sigma_T$ are calculated from equations (\ref{eq:q2dsameside}) or (\ref{eq:q2doppside}).  

As an example of using these formulae, we will calculate the effective transport quantities for the samples used in Tamura \textit{et al's} experiments.  In the first configuration (contacts on opposite sides), we find that the effective resistance is $1.43R_{z0}$, close to $R_{z0}$ as expected on account of the large electrical anisotropy.  The effective thermal conductance is $0.044K_{z0}$, which is substantially reduced on account of the weak thermal anisotropy and small current contact.  The effective thermal resistance is therefore increased much more than the effective electrical resistance.  The effective thermopower is $1.87S_z$, which is very close to $S_x$ as predicted from the Harman FOM.  Finally, the effective FOM is $55.3z_z$, which is consistent with an average value over the current contact (see figure \ref{fig:tempcut}).  

In the second configuration, we find the effective resistance, thermopower and thermal conductance to be $0.42 R_{z0}$, $1.89S_z$ and $0.047K_{z0}$.  It is more difficult to interpret the results from this configuration, since the `normal' results for large contacts cannot be obtained in this case.  The most important feature is that the effective resistance is not exponentially small, unlike the measured resistance on the top of the sample.  This is testament to the fact that the effective resistance corresponds to the physical resistance of the sample.  It is however substantially smaller than $R_{z0}$, simply because the electrical current is not forced to flow in the $z$-direction where the resistivity is largest.  The effective thermal conductivity is still substantially smaller than $K_{z0}$, while the effective thermopower is again very close to $S_x$.  The effective FOM is much larger in this configuration, with $z_{\mathrm{eff}} = 178z_z$, primarily due to the lower electrical resistance. 

We conclude that the effective electrical resistance and thermal conductance describe the ways in which electricity and heat flow through the sample, while the effective thermopower captures the thermoelectric processes.  All three quantities include the effects of eddy currents and the Bridgman effect in an appropriate way.  However, we have not yet shown that the effective FOM is in any way related to the efficiency of a thermoelectric device.  Suppose we used the sample configuration shown in figure \ref{fig:Sample4}(a) to create a simple thermoelectric generator.  Hot and cold reservoirs would be connected over the current contacts, generating a unform thermoelectric voltage between the current contacts which could then be applied to an external load.  Our analysis of the Harman method implies that due to the anisotropy the thermal resistance will be increased much more than the electrical resistance.  Further, the thermoelectric voltage will be increased due to the eddy currents and Bridgman effect.  We would therefore expect the thermoelectric performance to be improved compared to the situation where the reservoirs are connected across the full area of the sample.  Since the effective FOM of the sample in this configuration is increased relative to $z_z$, we conclude that the effective FOM is at least qualitatively related to the thermoelectric performance of the sample when viewed as a thermoelectric device.  

Although we have shown the effective FOM is enhanced compared to the intrinsic FOM in one direction, we have not yet compared it with the intrinsic figures of merit in the other coordinate directions.  For example, in the first configuration of Tamura \textit{et al's} experiment, we found that effective FOM was increased by $55.3$ compared to $z_z$.  However, the intrinsic FOM in the $x$-direction for this material is $z_x = (r^2s^2/k^2)z_z = 9025z_z$!  Therefore, although the effective FOM is substantially enhanced compared to the intrinsic FOM in one direction, it is still much smaller than the intrinsic FOM in the other.  Recall that if the current is passed into the sample in a particular direction through large contacts (and with current leads of zero thermopower), then the effective FOM is just the intrinsic FOM for that particular direction.  Therefore, one can only really claim that the thermoelectric performance is improved with small contacts if the effective FOM exceeds \textit{all} of the intrinsic figures of merit.  Otherwise, it would always be better to pass the current though large contacts in the direction with the highest intrinsic FOM.  We will explore this issue in the next section, by establishing some bounds on the effective FOM.   

\section{Bounds on the Effective Figure of Merit}\label{sec:bounds}  

The aim of this section is to establish some bounds on the effective FOM.  It is convenient for this analysis to set the lead thermopower $S_l$ equal to zero.  None of the results we present rely on this assumption, but the analysis becomes considerably simpler with it \cite{comment3}.  We will prove that in an anisotropic system with any contact configuration, the effective FOM is always bounded from above by the largest intrinsic FOM of the material.  Here, we will prove the result in the case of tetragonal symmetry and isotropic thermopower in order to illustrate the method.  This argument is then extended to include anisotropic thermopower in appendix \ref{sec:proof}.  The simplest way to prove this result is to return to equations (\ref{eq:bigchi}) and (\ref{eq:bigT}).  Restricting to tetragonal symmetry and isotropic thermopower, and also setting $S_l=0$, the ac voltage and temperature distributions may be written as
\begin{eqnarray}
\chi_{\mathrm{ac}} = \frac{I}{\sigma_z}\sum_{mnp}'\frac{A_{mnp}}{r^2\omega_{mn}^2 + \tp^2}\coss{\tm}{x}\cos \tn y \cos \tp z, \\
T = \frac{\Pi I}{\kappa_z}\sum_{mnp}'\frac{A_{mnp}}{k^2\omega_{mn}^2 + \tp^2}\coss{\tm}{x}\cos \tn y \cos \tp z.
\end{eqnarray}
The quantities $A_{mnp}$ contain all of the information about the boundary conditions associated with the current contacts.  For example, for current contacts on the face at $z=0$, we would have $A_{mnp} = (\gamma_{mn} - \nu_{mn})/cd^2$.  The proof does not require any assumptions about the positions or sizes of the current contacts, with all of the relevant information subsumed into $A_{mnp}$.  The important point is that the information about the current contacts enters both $\chi_{\mathrm{ac}}$ and $T$ in the \textit{same} way through $A_{mnp}$.

Using these expressions the rates of entropy production $\Sigma_J$ and $\Sigma_T$ may be derived.  The effective FOM is found to be  
\begin{equation}\label{eq:zzbound}
z_{\mathrm{eff}} = z_z\frac{\displaystyle{\sum_{mnp}'\frac{A^2_{mnp}}{k^2\omega_{mn}^2 + \tp^2}}}{\displaystyle{\sum_{mnp}'\frac{A^2_{mnp}}{r^2\omega_{mn}^2 + \tp^2}}}.
\end{equation}
We now compare corresponding terms (\textit{i.e.} those with the same values of $m$, $n$, and $p$) in the numerator and denominator.  Note that every term in the numerator and denominator of this expression is positive.  For $r > k$, it is clear that every term in the numerator is greater than or equal to the corresponding term in the denominator.  Equality holds for $\omega_{mn}=0$.  Therefore, the sum in the numerator is greater than or equal to the sum in the denominator.  Hence for $r > k$ it follows that $z_{\mathrm{eff}} \ge z_z$.  Conversely, for $r < k$ every term in the numerator is less than or equal to the corresponding term in the denominator, and $z_{\mathrm{eff}} \le z_z$.

The next step is to investigate the relative sizes of $z_{\mathrm{eff}}$ and $z_x$.  Noting that for isotropic thermopower $z_x = (r^2/k^2)z_z$, equation (\ref{eq:zzbound}) can be rewritten as 
\begin{equation}\label{eq:zxbound}
z_{\mathrm{eff}} = z_x\frac{\displaystyle{\sum_{mnp}'\frac{A^2_{mnp}}{\omega_{mn}^2 + \tp^2/k^2}}}{\displaystyle{\sum_{mnp}'\frac{A^2_{mnp}}{\omega_{mn}^2 + \tp^2/r^2}}}.
\end{equation}
The analysis above can now be repeated for $z_x$.  For $r > k$ every term in the numerator is less than or equal to the corresponding term in the denominator, and consequently $z_{\mathrm{eff}} \le z_x$.  Conversely, for $r < k$ every term in the numerator is greater than or equal to the corresponding term in the denominator, and so $z_{\mathrm{eff}} \ge z_x$.

A summary of the various cases is given below:
\begin{align}
&\mathrm{for} \ r/k > 1, \ z_z \le z_{\mathrm{eff}} \le z_x \nonumber; \\
&\mathrm{for} \ r/k < 1, \ z_x \le z_{\mathrm{eff}} \le z_z \nonumber; \\
&\mathrm{for} \ r/k = 1, \ z_z = z_{\mathrm{eff}} = z_x \nonumber; 
\end{align}  
where the final case trivially follows from equations (\ref{eq:zzbound}) and (\ref{eq:zxbound}).  We conclude that when the thermopower is isotropic the effective FOM is always less than or equal to the largest intrinsic FOM, in any contact configuration.

The extension of this proof to anisotropic thermopower may be found in appendix \ref{sec:proof}.  The analysis is very similar to that presented here, but with several additional complications.  For anisotropic thermopower the intrinsic figures of merit are related by $z_x = (r^2 s^2/k^2) z_z$.  The parameter which controls the relative sizes of $z_{\mathrm{eff}}$, $z_z$ and $z_x$ is $r|s|/k$.  The modulus signs are needed since $s=S_x/S_z$ could be either positive or negative.  The various cases are:
\begin{align}
&\mathrm{for} \ r|s|/k > 1, \ z_z \le z_{\mathrm{eff}} \le z_x \ \mathrm{or} \ z_{\mathrm{eff}} \le z_z \le z_x;  \nonumber \\
&\mathrm{for} \ r|s|/k < 1, \ z_x \le z_{\mathrm{eff}} \le z_z \ \mathrm{or} \ z_{\mathrm{eff}} \le z_x \le z_z;  \nonumber \\
&\mathrm{for} \ r|s|/k = 1, \ z_{\mathrm{eff}} \le z_z \ \mathrm{and} \ z_{\mathrm{eff}} \le z_x. \nonumber
\end{align}  
Again, it is clear that $\zeff$ is always less than or equal to the largest intrinsic FOM.  The added complication when the thermopower is anisotropic is that it is much harder to define the regime where each intrinsic FOM is larger than $\zeff$.  This regime partly depends on other factors such as the current contact locations and sizes.  It is also possible for $\zeff$ to be simultaneously smaller than both intrinsic figures of merit, which was not possible for isotropic thermopower.

In summary then, we have shown in this section that for a sample with tetragonal symmetry and an arbitrary arrangement of current contacts the effective FOM is always bounded from above by the largest intrinsic FOM of the material.  Therefore, if the maximum FOM is the only consideration, it is always better to pass the current through large contacts in the direction where the intrinsic FOM is highest.  It should be possible to extend this analysis beyond tetragonal symmetry to the most general case of anisotropy.  We also note that our results are somewhat similar to those of Bergman \textit{et al} \cite{Bergman}, who proved that for a composite thermoelectric material the overall FOM cannot exceed any of the individual figures of its constituents.    

\section{Conclusion}\label{sec:conc}

We have investigated the effects of small current contacts and anisotropy on the measurement of the FOM using the Harman method.  Our model captures the non-trivial effects associated with thermoelectric anisotropy in the form of thermoelectric eddy currents and the Bridgman effect.  This introduces a self-consistency between the temperature and voltage distributions.  Using our solutions, we investigated what would be measured in the Harman method with small current contacts.  We found that the Harman FOM, $z_H$, is generally very different from the intrinsic FOM of the sample, except in a few special cases.  When the thermopower is isotropic, the Harman FOM may still be interpreted in terms of a measured electrical resistance and thermal conductance.  This interpretation is no-longer possible when the thermopower is anisotropic, primarily due to the breakdown of the relationship between the dc, ac and thermoelectric voltages.  We also investigated how geometrical factors can also change the Harman FOM.  The small contact version of the conventional Harman method (current and voltage contacts on opposite sides of the sample) is particularly sensitive to the anisotropy, geometry and current contact sizes and locations.  We believe that this is the main reason reason for the poor quantitative agreement between the theory and the Harman FOM measured in the first configuration of Tamura \textit{et al's} experiments, although we cannot rule-out some other non-trivial effects not captured by this theory.  A key test of the theory would be to measure the Harman FOM in one of configurations 1, 2 or 4 of table \ref{tab:same} in the thick sample limit.  The Harman FOM in these cases depends on all of the anisotropy parameters, but is only weakly dependent on the contact sizes and locations.  

We also extended the analysis of van der Pauw for thermoelectric measurements thin samples of arbitrary shape. We showed that there exist van der Pauw-type formulae relating particular linear combinations of the potential and temperature differences measured between points on the edge of the sample. 

The work on thermoelectric measurements in anisotropic systems is not yet complete.  Further analysis is required to determine the effects of heat losses, which were neglected in our calculations.  Overall though, the main message from this part of the work is that great care should be taken in interpreting the FOM measured using the Harman method, particularly in anisotropic systems.   

In the second part of the paper we explored whether the increased Harman FOM measured in certain sample configurations actually indicates an improvement in the thermoelectric performance.  We argued that an effective FOM could be defined from the rates of entropy production occurring in the sample in the Harman set-up, $z_{\mathrm{eff}} = \Sigma_T/\Sigma_J$.  Unlike the Harman FOM, this quantity takes a single value for a sample with a given configuration of current contacts.  We identified two effects of small current contacts and anisotropy which can cause the effective FOM to be enhanced.  First, the flow of heat can confined relative to the flow of electricity, leading to the effective thermal resistance being increased more than the effective electrical resistance.  Second, the effective thermoelectric power can also be enhanced by the Bridgman effect combined with thermoelectric eddy currents.  However, we also proved that for a sample with small current contacts and whose transport coefficients have tetragonal symmetry, the effective FOM is always bounded from above by the largest intrinsic FOM of the material.  

Despite this fact, the use of anisotropy to improve thermoelectric performance in the manner indicated in this work may still be valid.  It is important to remember that the intrinsic FOM only determines the efficiency of a thermoelectric device under certain assumptions (as described in the introduction), and is generally used as a guide for selecting materials as potential thermoelectrics.  Other factors such as the mechanical properties of thermoelectric materials or manufacturing issues are also important considerations in developing thermoelectric devices.  The primary reason for the enhanced effective FOM in an anisotropic system with small current contacts is that heat and electricity can be made to flow along different paths.  This ability to independently control the flows of heat and electricity in anisotropic systems is one of the main findings of this work, and may itself have useful applications.

\section*{Acknowledgements}
We would like to thank I. Terasaki for introducing us to this problem, and for useful insights and advice. We also thank N. E. Hussey for helpful discussions.  T. W. S. acknowledges support from the EPSRC.

\begin{appendix}

\section{Solution for Current Contacts on the Same Face of the Sample}\label{sec:same}

In this configuration, both contacts are located on the face at $z=0$, and hence we set $z_1=0=z_2$ in equations (\ref{eq:bigchi}) and (\ref{eq:bigT}).  At this stage, we can also remove the restriction to point contacts.  To do this, we `broaden' the delta-functions in the $x$ and $y$-directions into top-hat functions of width $d$; $\delta(x-x_1) \rightarrow (1/d)\Theta(d/2-|x-x_1|)$ \textit{etc}.  The point contacts results are recovered in the limit $d\rightarrow 0$.  The Fourier representation of the contact located at $(x_1,y_1,z_1)$ is given by
\begin{multline}
\frac{1}{d^2}\Theta(d/2-|x-x_1|)\Theta(d/2-|y-y_1|) = \\ 
\frac{1}{d^2}\sum_{mn}\nu_{mn}\coss{\tilde m}{x}\coss{\tilde n}{y};
\end{multline}
\begin{align}\label{eq:fourierco}
\nu_{00} &= \frac{d^2}{ab}, \nonumber \\
\nu_{m0} &= \displaystyle{\frac{d}{b}\frac{2}{(m \pi)}\sin \frac{m \pi d}{2a}\coss{\tilde m}{x_1}}, \nonumber \\
\nu_{0n} &= \displaystyle{\frac{d}{a}\frac{2}{(n \pi)}\sin \frac{n \pi d}{2b}\coss{\tilde n}{y_1}}, \nonumber \\
\nu_{mn} &= \displaystyle{\frac{4}{(m \pi)(n \pi)}\sin \frac{m \pi d}{2a}\sin \frac{n \pi d}{2b}\coss{\tilde m}{x_1}\coss{\tilde n}{y_1}}.
\end{align}
The second contact, located at $(x_2,y_2,z_2)$ has a similar Fourier representation with coefficients $\gamma_{mn}$, which may be obtained from $\nu_{mn}$ with the replacement $1 \rightarrow 2$.

A considerable simplification of (\ref{eq:bigchi}) and (\ref{eq:bigT}) can be obtained by performing the sums over $p$.   To do this, we define the following anisotropy parameters
\begin{align}
u_{mn}^2 &= \frac{\sigma_x}{\sigma_z}\tm^2 + \frac{\sigma_y}{\sigma_z}\tn^2 = r_{xz}^2\tm^2 + r_{yz}^2\tn^2, \\
v_{mn}^2 &= \frac{\sigma_x}{\sigma_z}\frac{S_x}{S_z}\tm^2 + \frac{\sigma_y}{\sigma_z}\frac{S_y}{S_z}\tn^2, \nonumber \\
&= r_{xz}^2s_{xz}\tm^2 + r_{yz}^2s_{yz}\tn^2, \\
w_{mn}^2 &= \Big (\frac{\sigma_x}{\sigma_z}\frac{S_x^2}{S_z^2} + \frac{\kappa_x}{\kappa_z}z_z^{-1} \Big )\tm^2 + \Big (\frac{\sigma_y}{\sigma_z}\frac{S_y^2}{S_z^2} + \frac{\kappa_y}{\kappa_z}z_z^{-1} \Big )\tn^2 \nonumber \\
&= (r_{xz}^2s_{xz}^2+k_{xz}^2z_z^{-1})\tm^2 + (r_{yz}^2s_{yz}^2 + k_{yz}^2z_z^{-1})\tn^2. 
\end{align}
We have also defined the intrinsic FOM in the $z$-direction, $z_z = \bar T S_z^2 \sigma_z / \kappa_z$.  In terms of these parameters, $\chi^s$ and $T^s$ may be written as
\begin{widetext}
\begin{align}
\left(
\begin{array}{c}
\chi^s \\
S_z T^s
\end{array}
\right) &=
\frac{I}{\sigma_z d^2 c} \sum_{mnp}'\frac{(\gamma_{mn}-\nu_{mn})\coss{\tm}{x}\coss{\tn}{y}\coss{\tp}{z}}{\big (\tp^2+u_{mn}^2 \big )\big ((1+z_z^{-1})\tp^2+w_{mn}^2 \big) - \big (\tp^2+v_{mn}^2 \big )^2}
\left(
\begin{array}{c}
w_{mn}^2-s_l v_{mn}^2 - (s_l - 1 - z_z^{-1})\tp^2 \\
v_{mn}^2-s_lu_{mn}^2 - (s_l-1)\tp^2
\end{array}
\right) \nonumber \\
\label{eq:intsameside}
&= \frac{I z_z}{\sigma_z d^2 c} \sum_{mnp}'\frac{(\gamma_{mn}-\nu_{mn})}{\lambda_2^2-\lambda_1^2}\coss{\tm}{x}\coss{\tn}{y}\sum_{i=1,2}(-1)^{i+1}
\left(
\begin{array}{c}
w_{mn}^2-s_l v_{mn}^2 + (s_l - 1 - z_z^{-1})\lambda_i^2 \\
v_{mn}^2-s_lu_{mn}^2 + (s_l-1)\lambda_i^2
\end{array}
\right)
\frac{\coss{\tp}{z}}{\tp^2+\lambda_i^2}, \\
\label{eq:lambda}
\lambda_i^2 &= \frac{(z_z^{-1}+1)u_{mn}^2 - 2v_{mn}^2 + w_{mn}^2+(-1)^{i+1}\sqrt{\big(z_z^{-1}+1)u_{mn}^2 - 2v_{mn}^2 + w_{mn}^2 \big)^2-4z_z^{-1}(u_{mn}^2w_{mn}^2-v_{mn}^4)}}{2z_z^{-1}},
\end{align}
where we have also defined a dimensionless thermopower of the current leads, $s_l = S_l/S_z$.  In going from the first to the second line, the denominator has been factorised into the form $(\tp^2 + \lambda_1^2)(\tp^2+\lambda_2^2)$, and then the expression split into partial fractions. This allows the sum over $p$ to be performed, giving the final result 
\begin{align}\label{eq:sameside}
\left(
\begin{array}{c}
\chi^s \\
S_z T^s
\end{array}
\right) &=
\frac{I z_z}{\sigma_z d^2} \sum_{mn}'\sum_{i=1,2}\frac{(-1)^{i+1}}{\lambda_2^2-\lambda_1^2}
\left(
\begin{array}{c}
w_{mn}^2-s_l v_{mn}^2 + (s_l - 1 - z_z^{-1})\lambda_i^2 \\
v_{mn}^2-s_lu_{mn}^2 + (s_l-1)\lambda_i^2
\end{array}
\right)
\Lambda_{mn}(x,y,z;\lambda_i),
\end{align}
\end{widetext}
where 
\begin{multline}\label{eq:Lambda}
\Lambda_{mn}(x,y,z;\lambda_i) = \\ 
(\gamma_{mn}-\nu_{mn})\frac{\cosh \lambda_i (z-c)}{\lambda_i \sinh \lambda_i c}\coss{\tm}{x}\coss{\tn}{y}.
\end{multline}
Note that in the original representation (\textit{i.e.} before summing over $p$) all terms with $m=n=0$ vanished independently of $p$.  Hence, the term with $m=n=0$ is excluded from the sum in equation (\ref{eq:sameside}).  

Note that in having performed the sum over $p$, the boundary conditions satisfied by the functions $\chi$ and $\eta$ have changed.  The solutions now satisfy homogeneous versions of equations (\ref{eq:eleccont2}) and (\ref{eq:heatcont2}) \textit{i.e.} without the source/sink terms on the right-hand side.  The boundary conditions on the faces at $x=0,a$, $y=0,b$ and $z=c$ remain unchanged, but the boundary conditions on the face at $z=0$ are altered to read
\begin{align}\label{eq:elecheatbcs}  
j_z|_{z=0} = \frac{I}{d^2}  \Big ( &\Theta (d/2 -|x-x_1|)\Theta (d/2 -|y-y_1|) \nonumber \\
 - &\Theta (d/2 -|x-x_2|)\Theta (d/2 -|y-y_2|) \Big ), \nonumber \\
j_{Qz}|_{z=0} = \frac{\Pi_lI}{d^2} \Big ( &\Theta (d/2 -|x-x_1|)\Theta (d/2 -|y-y_1|) \nonumber \\
 - &\Theta (d/2 -|x-x_2|)\Theta (d/2 -|y-y_2|) \Big ). \nonumber 
\end{align}
These boundary conditions correspond to a constant flux of electrical current (and Peltier heat) into the contact at $(x_1,y_1,0)$ and out of the contact at $(x_2,y_2,0)$.  Indeed, if we had started with this boundary condition initially, along with the homogeneous forms of equations (\ref{eq:eleccont2}) and (\ref{eq:heatcont2}), then equations (\ref{eq:sameside}) would have been obtained directly.

\section{Solution for Current Contacts on Opposite Faces of the Sample}\label{sec:opp}

In this configuration, one contact is placed on the face at $z=0$, while the other is placed on the face at $z=c$.  Hence, we simply set $z_1=0$ and $z_2=c$ in equations (\ref{eq:bigchi}) and (\ref{eq:bigT}). The term proportional to $\gamma_{mn}$ now acquires a prefactor of $(-1)^p$.  This leads to an important difference with the previous case, namely that not all of the terms with $m=n=0$ vanish.  The terms with $m=n=0$ are given by
\begin{multline}
\frac{I}{\sigma_z abc}\left(
\begin{array}{c}
1-\sigma_zS_z(\Pi_l-\Pi_z)/\kappa_z\\
-\sigma_zS_z(\Pi_l-\Pi_z)/\kappa_z
\end{array}
\right) \\
\times \sum_p \big((-1)^p-1)\frac{\coss{\tp}{z}}{\tp^2} \\
= \frac{Ic}{\sigma_zab}\Big(\frac{z}{c}-\frac{1}{2}\Big)
\left(
\begin{array}{c}
1-\sigma_zS_z(\Pi_l-\Pi_z)/\kappa_z\\
-\sigma_zS_z(\Pi_l-\Pi_z)/\kappa_z
\end{array}
\right).
\end{multline}
Rewriting the remaining terms in the same way as before, and performing the summation over $p$, we find the final result for contacts on opposite faces
\begin{widetext}
\begin{multline}\label{eq:oppside}
\left(
\begin{array}{c}
\chi^o \\
S_z T^o
\end{array}
\right) =
\frac{Ic}{\sigma_zab}\Big(\frac{z}{c}-\frac{1}{2}\Big)
\left(
\begin{array}{c}
1-\sigma_zS_z(\Pi_l-\Pi_z)/\kappa_z\\
-\sigma_zS_z(\Pi_l-\Pi_z)/\kappa_z
\end{array}
\right) + \\
\frac{Iz_z}{\sigma_z d^2} \sum_{mn}'\sum_{i=1,2}\frac{(-1)^{i+1}}{\lambda_2^2-\lambda_1^2}
\left(
\begin{array}{c}
w_{mn}^2-s_l v_{mn}^2 + (s_l - 1 - z_z^{-1})\lambda_i^2 \\
v_{mn}^2-s_lu_{mn}^2 + (s_l-1)\lambda_i^2
\end{array}
\right)
\Omega_{mn}(x,y,z;\lambda_i)
\end{multline}
\end{widetext}
where
\begin{multline}
\Omega_{mn}(x,y,z;\lambda_i) = \\
\frac{\gamma_{mn} \cosh \lambda_i z - \nu_{mn} \cosh \lambda_i (z-c)}{\lambda_i \sinh \lambda_i c}\coss{\tm}{x}\coss{\tn}{y}.
\end{multline}

\section{Entropy Production and Contact Averaging}\label{sec:ent}

Here we will prove the results quoted in the main text, which relate the rates of entropy production to contact averaged temperature and potential differences.  We will prove the results for isotropic thermopower.  In addition, we will use a sample configuration with the finite-size current contacts on the same side of the sample at $z=0$.  This choice is arbitrary, but the analysis is simpler to follow with a definite example.  Given these restrictions, the voltage and temperature distributions in the sample satisfy 
\begin{multline}\label{eq:entiso}       
\sigma_x \partwo{\chi}{x} + \sigma_y \partwo{\chi}{y} + \sigma_z \partwo{\chi}z  \\
- \sigma_x S \partwo{T}{x} - \sigma_{y} S \partwo{T}{y} - \sigma_z S \partwo{T}{z} = I\Delta(x,y,z),
\end{multline}
\begin{equation}
\displaystyle{-\kappa_x\partwo{T}{x} - \kappa_y\partwo{T}{y} - \kappa_z\partwo{T}{z} = (\Pi_l-\Pi)I\Delta(x,y,z)},
\end{equation}
with
\begin{multline}
\Delta(x,y,z) = \frac{1}{d^2}\big(\Theta(d/2-|x-x_1|)\Theta(d/2-|y-y_1|) \\
-\Theta(d/2-|x-x_2|)\Theta(d/2-|y-y_2|)\big)\delta(z).
\end{multline}
Consider the rate of entropy production due to Joule heating, $\Sigma_J$.  Substituting for $j_i$ using equation (\ref{eq:anisoelec}), integrating by parts, and using equation (\ref{eq:entiso}) provides 
\begin{align}
\bar T \Sigma_J &= \int dV \Big ( \frac{j_x^2}{\sigma_x} + \frac{j_y^2}{\sigma_y} + \frac{j_z^2}{\sigma_z} \Big )\nonumber \\
&=-\int dV (\chi - ST)I\Delta(x,y,z)
\end{align}
All surface terms vanish due to the insulating boundary conditions on the surfaces of the sample.  The quantity $\chi-ST$ is recognised as the ac voltage, $\chi_{\mathrm{ac}}$.  Substituting for $\Delta(x,y,z)$, this expression may be re-written as
\begin{multline}
\bar T \Sigma_J = I\bigg [\frac{1}{d^2} \int_{x_2-d/2}^{x_2+d/2}dx\int_{y_2-d/2}^{y_2+d/2}dy \chi_{\mathrm{ac}}(x,y,0) \\ 
-\frac{1}{d^2} \int_{x_1-d/2}^{x_1+d/2}dx\int_{y_1-d/2}^{y_1+d/2}dy\chi_{\mathrm{ac}}(x,y,0) \bigg ] = I\langle V_{\mathrm{ac}} \rangle .
\end{multline}
The quantity in square brackets is just the average ac potential difference between the current contacts.  Hence the total rate of Joule heating in the sample, $\bar T \Sigma_J$ is related to the average power supplied by the external current source, $I\langle V_{\mathrm{ac}} \rangle$.  

Similarly, it can be shown that the rate of entropy production due to heat flow, $\Sigma_T$, is related to the contact averaged temperature difference:
\begin{align}
\bar T \Sigma_T &= \int dV \Big ( \kappa_x\Big(\parone{T}{x}\Big)^2 + \kappa_y\Big(\parone{T}{y}\Big)^2 + \kappa_z\Big(\parone{T}{z}\Big)^2 \Big ) \nonumber \\
&= -I(S_l-S)\langle T \rangle 
\end{align}
Hence the total rate of entropy production is related to the contact averaged dc potential difference, $\bar T \Sigma = I\langle V_{\mathrm{dc}} \rangle = I(\langle V_{\mathrm{ac}} \rangle - (S_l-S)\langle \Delta T \rangle)$.

\section{Bounding the effective figure of merit; Anisotropic Thermopower}\label{sec:proof}

We will prove that in the case of tetragonal symmetry and anisotropic thermopower the effective FOM is always bounded from above by the largest intrinsic FOM of the material.  Using equations (\ref{eq:bigchi}) and (\ref{eq:bigT}) and setting $S_l=0$, the effective FOM is found to be
\begin{widetext}
\begin{equation}\label{eq:zsbound}
z_{\mathrm{eff}} = z_z\frac{\displaystyle{\sum_{mnp}'\frac{A^2_{mnp}}{D_{mnp}^2}\big(k^2\omega_{mn}^2 + \tp^2\big ) \big (r^2 s \omega_{mn}^2+\tp^2 \big )^2}}{\displaystyle{\sum_{mnp}'\frac{A^2_{mnp}}{D_{mnp}^2}\Big [ r^2\omega_{mn}^2\Big(k^2\omega_{mn}^2 + \big (1+(1-s)z_z \big )\tp^2 \Big )^2 + \tp^2 \Big ( \big (k^2-r^2z_zs(1-s) \big )\omega_{mn}^2 + \tp^2 \Big )^2 \Big ]}},
\end{equation}
where $D_{mnp} = (r^2\omega_{mn}^2 + \tp^2)(k^2\omega_{mn}^2 + \tp^2) + r^2z_z(1-s)^2\omega_{mn}^2\tp^2$.  The form of equation (\ref{eq:zsbound}) indicates that $\zeff$ is always positive.  Expanding the numerator and denominator, this equation has the form
\begin{equation}\label{eq:zsbound2}  
z_{\mathrm{eff}} = z_z\frac{\displaystyle{\sum_{mnp}'\frac{A^2_{mnp}}{D_{mnp}^2} \big (n_1 \omega_{mn}^6 + n_2 \omega_{mn}^4 \tp ^2 + n_3 \omega_{mn}^2 \tp^4 + n_4 \tp^6 \big )}}{\displaystyle{\sum_{mnp}'\frac{A^2_{mnp}}{D_{mnp}^2} \big (d_1 \omega_{mn}^6 + d_2 \omega_{mn}^4 \tp ^2 + d_3 \omega_{mn}^2 \tp^4 + d_4 \tp^6 \big )}} = z_x\frac{\displaystyle{\sum_{mnp}'\frac{A^2_{mnp}}{D_{mnp}^2} \big (n_1' \omega_{mn}^6 + n_2' \omega_{mn}^4 \tp ^2 + n_3' \omega_{mn}^2 \tp^4 + n_4' \tp^6 \big )}}{\displaystyle{\sum_{mnp}'\frac{A^2_{mnp}}{D_{mnp}^2} \big (d_1 \omega_{mn}^6 + d_2 \omega_{mn}^4 \tp ^2 + d_3 \omega_{mn}^2 \tp^4 + d_4 \tp^6 \big )}},
\end{equation}
\end{widetext}
We begin by establishing the regimes in which $\zeff$ is larger, smaller or equal to $z_z$.  To do this we consider the relative sizes of the coefficients of like powers of $\omega_{mn}$ and $\tp$ in the numerator and denominator of the first equality in equation (\ref{eq:zsbound2}) \textit{i.e.} $n_i$ and $d_i$.  The second equality will be used later to establish the size of $z_{\mathrm{eff}}$ relative to $z_x$.  The factors of $A_{mnp}^2/D_{mnp}^2$ may be ignored since they appear in both the numerator and denominator.  The coefficients $d_i$ of the terms in the denominator are all positive, as are $n_1$ and $n_4$.  The coefficients $n_2$ and $n_3$ can be negative for $s<0$, which occurs if the intrinsic thermopowers have opposite signs.  If we can show that $n_i/d_i \ge 1$ for all $i$, then the numerator of the first equality of equation (\ref{eq:zsbound2}) is definitely greater than or equal to the denominator, and $\zeff \ge z_z$.  Similarly, if we can show that $n_i/d_i \le 1$ for all $i$, then the numerator of equation (\ref{eq:zsbound2}) is definitely less than or equal to the denominator, and $\zeff \le z_z$.  This is essentially the strategy we will use in this proof.

\begin{figure}[h!]  
\subfigure[\ $\mathcal O(\omega_{mn}^6)$ and $\mathcal O(\tp^6)$]{\includegraphics[width=0.45\textwidth]{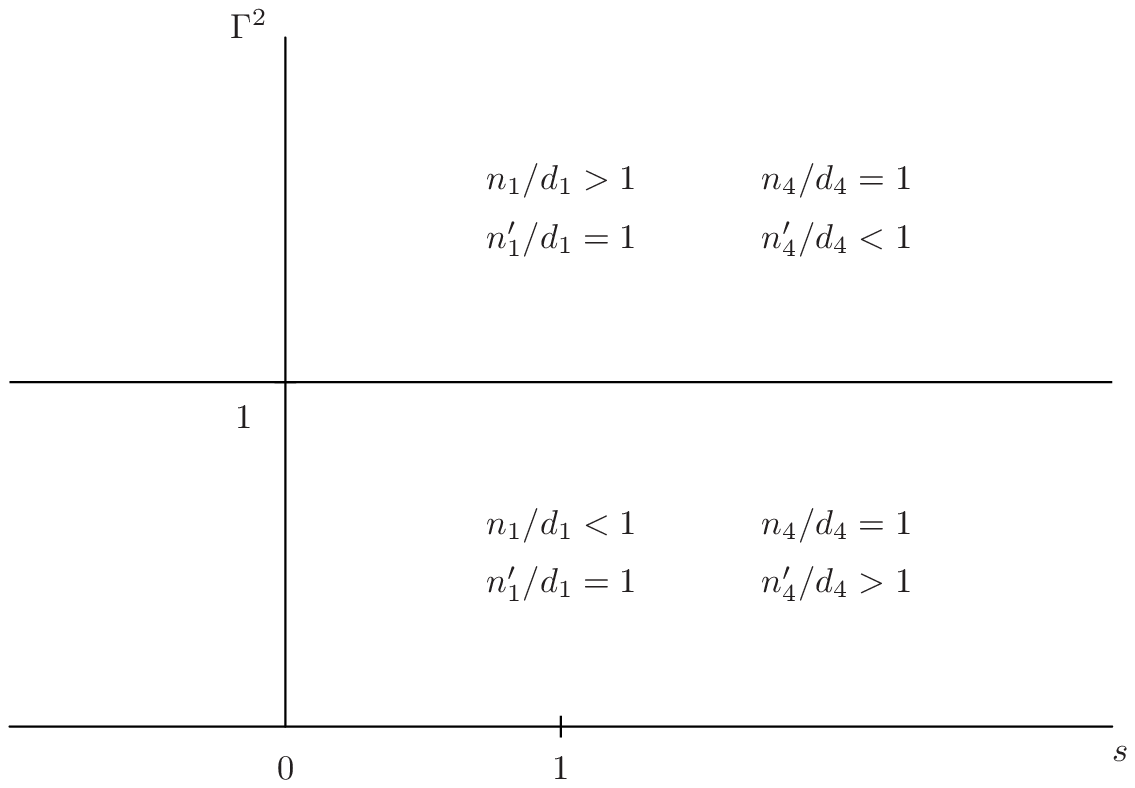}} \\
\subfigure[\ $\mathcal O(\omega_{mn}^4 \tp^2)$]{\includegraphics[width=0.45\textwidth]{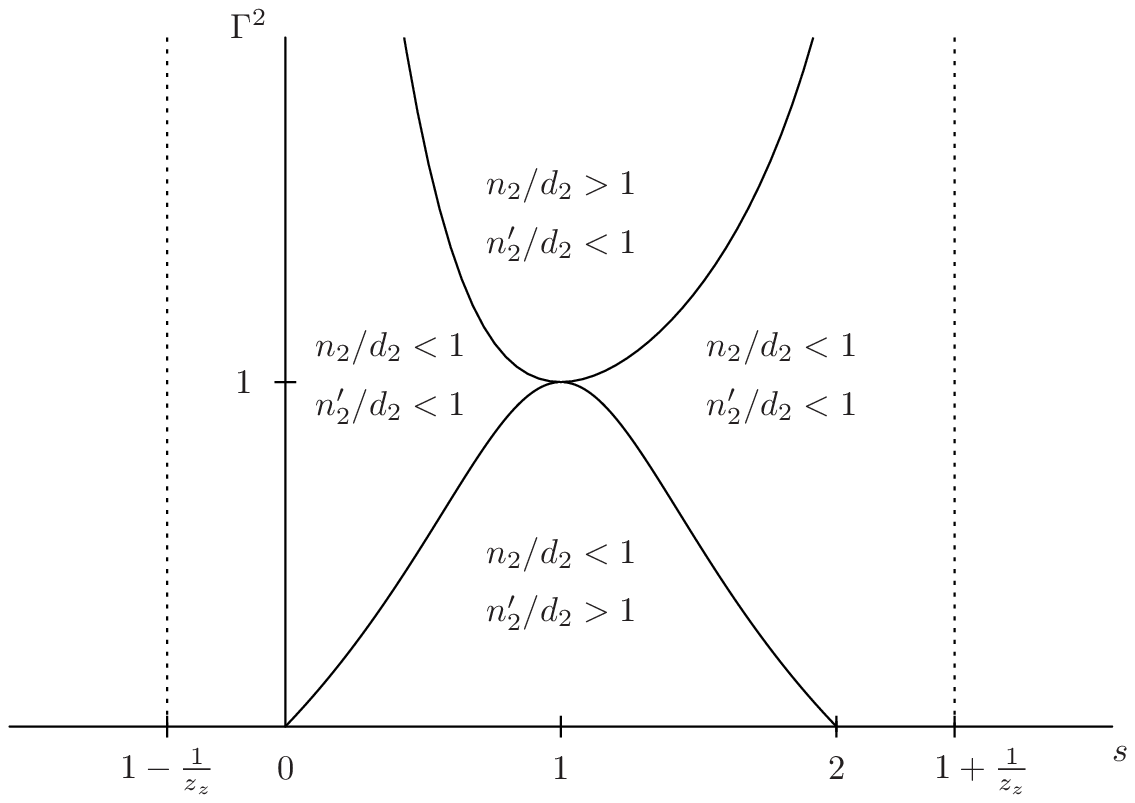}}     \\   
\subfigure[\ $\mathcal O(\omega_{mn}^2 \tp^4)$]{\includegraphics[width=0.45\textwidth]{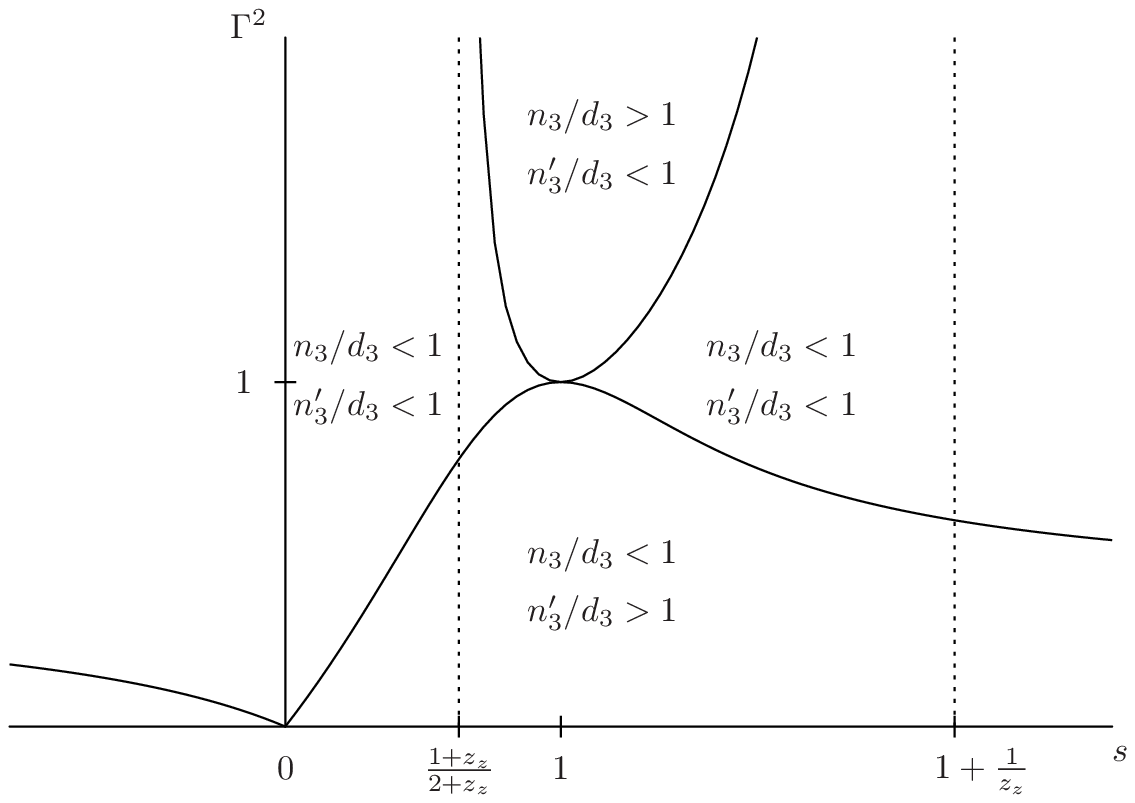}}
\caption{Plots indicating the relative sizes of the coefficients appearing in the expression for $\zeff$.  Solid lines are curves where $n_i/d_i=1$, while dotted lines indicate asymptotes to these curves.  (a) Coefficients of terms of $\mathcal O(\omega_{mn}^6)$ and  $\mathcal O(\tp^6)$;  (b) Coefficients of terms of  $\mathcal O(\omega_{mn}^4 \tp^2)$;  (c) Coefficients of terms of  $\mathcal O(\omega_{mn}^2 \tp^4)$. }
\label{fig:bounds}
\end{figure}

Taking first $\mathcal O(\omega_{mn}^6)$, we have $n_1/d_1 = \Gamma^2$, where $\Gamma^2 = r^2s^2/k^2$.  Hence it is clear that if $\Gamma^2 \ge 1$, $n_1/d_1 \ge 1$, while for $\Gamma^2 \le 1$, $n_1/d_1 \le 1$. The terms of $\mathcal O(\tp^6)$ are also easily bounded.  We find that $n_4/d_4 = 1$, independently of the value of $\Gamma^2$.  In figure \ref{fig:bounds}(a) we indicate the behaviour of $n_1/d_1$ and $n_4/d_4$ in the $\Gamma^2$-$s$ plane.  The reason for doing this will become clear shortly. 

Next we deal with the terms of $\mathcal O(\omega_{mn}^4 \tp^2)$:
\begin{equation}
\frac{n_2}{d_2} = \frac{\Gamma^4+2\Gamma^2s}{s^2+2\Gamma^2 + 2\Gamma^2z_z(1-s)^2 + \Gamma^4z_z^2(1-s)^2},
\end{equation} 
Unlike the previous two terms, $n_2/d_2$ is not simply a function of $\Gamma^2$, but also depends explicitly on $s$ and $z_z$.  Noting that $\Gamma^2$ is necessarily positive, the boundary where $n_2/d_2=1$ is given by
\begin{multline}\label{eq:n2d2}
\Gamma^2 = \frac{1-s+z_z(1-s)^2}{1-z_z^2(1-s)^2} \\
+ \frac{\sqrt{\big(1-s+z_z(1-s)^2 \big )^2 + s^2 \big (1-z_z^2(1-s)^2 \big )}}{1-z_z^2(1-s)^2}
\end{multline}
In figure \ref{fig:bounds}(b) we plot this expression as a function of $s$ for a fixed value of $z_z$.  The positive part of this function is contained within the two asymptotes at $z=1-1/z_z$ and $1+1/z_z$.  It can be shown that the minimum value this function is $\Gamma^2=1$ and occurs at $s=1$ independently of the value of $z_z$.  For $\Gamma^2$ lying above this line we have $n_2/d_2 > 1$, while for $\Gamma^2$ lying below the line, $n_2/d_2 < 1$.

The terms of $\mathcal O(\omega_{mn}^2 \tp^4)$ are analysed in exactly the same way:
\begin{equation}
\frac{n_3}{d_3} = \frac{s^2+2\Gamma^2s}{2s^2+\Gamma^2 + 2\Gamma^2z_z(1-s)^2 + \Gamma^2z_z^2(1-s)^2},
\end{equation} 
The boundary where $n_3/d_3=1$ is given by
\begin{equation}\label{eq:n3d3}
\Gamma^2 = \frac{s^2}{2s-1-2z_z(1-s)^2-z_z(1-s)^2} 
\end{equation}
This function is plotted in figure \ref{fig:bounds}(c).  The positive part of this function is contained within the two asymptotes at $z=(1+z_z)/(2+z_z)$ and $1+1/z_z$.  As before, it can be shown that the minimum value of this function is $\Gamma^2=1$ and occurs at $s=1$ independently of the value of $z_z$.  For $\Gamma^2$ lying above this line we have $n_3/d_3 > 1$, while for $\Gamma^2$ lying below the line, $n_3/d_3 < 1$.

We can now establish the relative sizes of $z_z$ and $\zeff$.  The most important regime to identify is where $\zeff$ is \textit{definitely} less than or equal to $z_z$.  The condition for this to be true is that $n_i/d_i \le 1$ for all $i$.  An examination of figures \ref{fig:bounds}(a)-(c) reveals that this condition is met for $\Gamma^2 \le 1$, independently of $s$.  There is also region where $n_i/d_i \ge 1$ for all $i$, meaning $z_{\mathrm{eff}}$ is definitely greater than or equal to $z_z$.  Finally, there is a region where definite statements cannot be made about the relative sizes of $\zeff$ and $z_z$.  In this region, $n_i/d_i \ge 1$ for one or more $i$, and $n_i/d_i \le 1$ for the others.  Here, the overall size of $z_{\mathrm{eff}}$ relative to $z_z$ depends on a competition between all of the terms in the sum, and will also depend on the quantities $A_{mnp}$ which encode the current contact positions and geometry.  

Having established the bounds on the effective FOM relative to $z_z$, we now need to repeat the entire procedure for $z_x$.  For anisotropic thermopower the relationship between $z_z$ and $z_x$ is $z_x = (r^2s^2/k^2)z_z = \Gamma^2 z_z$.  We now use the second equality in equation (\ref{eq:zsbound2}), where $n_i' = n_i/\Gamma^2$.  At $\mathcal O (\omega_{mn}^6)$ we find $n_1'/d_1 = 1$ independently of $\Gamma^2$, while at $\mathcal O(\tp^6)$ we have $n_4'/d_4 = 1/\Gamma^2$.  The various cases for $n_1'/d_1$ and $n_4'/d_4$ are shown in figure \ref{fig:bounds}(a).  For $\mathcal O(\omega_{mn}^4\tp^2)$ the curve on which $n_2'/d_2=1$ is given by     
\begin{multline}\label{eq:n2primed2}
\Gamma^2 = -\frac{1 + 2z_z(1-s)^2}{2z_z^2(1-s)^2} \\
+ \frac{\sqrt{\big(1 + 2z_z(1-s)^2 \big )^2 + 4s z_z^2(2-s)(1-s)^2}}{2z_z^2(1-s)^2},
\end{multline}     
which is sketched in figure \ref{fig:bounds}(b).  This function is only positive for $0 \le s \le 2$, and takes its maximum value at $s=1$, $\Gamma^2=1$ independently of the value of $z_z$.  Finally, for $\mathcal O(\omega_{mn}^2 \tp^4)$ the curve on which $n_3'/d_3=1$ is given by
\begin{multline}\label{eq:n3primed3}
\Gamma^2 = \frac{s-s^2}{1+2z_z(1-s)^2+z_z^2(1-s)^2} \\
+ \frac{\sqrt{\big(s-s^2 \big )^2 + s^2\big (1 + 2z_z(1-s)^2+z_z^2(1-s)^2\big )}}{1+2z_z(1-s)^2+z_z^2(1-s)^2},
\end{multline}     
which is plotted in figure \ref{fig:bounds}(c).  This function is positive for all $s$, and also takes its maximum value at $s=1$, $\Gamma^2=1$. 

We can now set bounds on the size of $\zeff$ relative to $z_x$.  Again, the most important region to define is where $\zeff$ is definitely less than or equal to $z_x$.  The condition for this is that $n_i'/d_i \le 1$ for all $i$.  A consideration of figures \ref{fig:bounds}(a)-(c) reveals that this condition is met for $\Gamma^2 \ge 1$, independently of $s$.  As for $z_z$, there is also a region where $\zeff$ is definitely greater than or equal to $z_x$, and another where definite statements cannot be made. 

We now have all of the information required to prove the result that the effective FOM is bounded from above by the largest intrinsic FOM.  The crucial statement is that the $\Gamma^2$-$s$ plane may be divided up into two regions.  For $\Gamma^2 \le 1$, $\zeff$ is \textit{definitely} less than or equal to $z_z$, independently of $s$.  For $\Gamma^2 \ge 1$, $\zeff$ is \textit{definitely} less than or equal to $z_x$, again independently of $s$.  This is sufficient to prove the result.  At any point in the $\Gamma^2$-$s$ plane, we can say with certainty that $\zeff$ is less than or equal to the largest intrinsic FOM.  The fact that the relationship between $\zeff$ and the other intrinsic FOM may be indeterminate is irrelevant.  Finally, we note that this proof fully reproduces the results of section \ref{sec:bounds} for isotropic thermopower, which follow by restricting to the line $s=1$ in the $\Gamma^2$-$s$ plane.  

\end{appendix}


\begin{thebibliography}{99}
\bibitem{comment2}
We define the delta function such that if the point $x_1$ lies on the boundary at $x=0$ $(x=a)$, the total weight of the delta-function is taken to lie above (below) $x = 0$ ($x=a$).  This will ensure that the total current flowing through the sample is $I$, without the need to adjust the source/sink strength for contacts on the faces/edges/corners.
\bibitem{comment3}
If the thermopower of the current leads is included, then the correct quantities to compare with the effective FOM are not the intrinsic figures of merit $z_x$, $z_y$ and $z_z$. Instead, the appropriate quantities for comparison are the `relative' figures of merit $\bar T \sigma_x(S_l-S_x)^2/\kappa_x$, $\bar T \sigma_y(S_l-S_y)^2/\kappa_y$ and $\bar T \sigma_z(S_l-S_z)^2/\kappa_z$.  It is this complication that we wish to avoid in the analysis.
\end{thebibliography}
\end{document}